\documentclass[11pt,a4paper]{article}
\usepackage{jheppub,xcolor}
\pdfoutput=1
\usepackage[utf8]{inputenc}
\graphicspath{{./figs/}}

\usepackage{amssymb}
\usepackage{dcolumn}	
\usepackage{bm}			
\usepackage{bbm} 		
\usepackage{multirow}
\usepackage{slashed}
\usepackage{blkarray}
\usepackage{booktabs}
\usepackage{makecell}
\usepackage{longtable}

\definecolor{Red}{rgb}{1.,0.,0.}

\newcommand{\lag}{\mathcal{L}}
\newcommand{\mcM}{\mathcal{M}}
\newcommand{\mcO}{\mathcal{O}}

\newcommand{\rtsha}{\sqrt{\hat{s}}}

\newcommand{\mz}{M_Z}
\newcommand{\mh}{M_h}

\newcommand{\Ohq}{\mcO_{\varphi q}^{(1)}}
\newcommand{\Ohqt}{\mcO_{\varphi q}^{(3)}}
\newcommand{\Ohu}{\mcO_{\varphi u}}
\newcommand{\Ohd}{\mcO_{\varphi d}}

\newcommand{\chq}{c_{\varphi q}^{(1)}}
\newcommand{\chqt}{c_{\varphi q}^{(3)}}
\newcommand{\chu}{c_{\varphi u}}
\newcommand{\chd}{c_{\varphi d}}

\newcommand{\cfu}{c_{\varphi u}}
\newcommand{\cfd}{c_{\varphi d}}
\newcommand{\cfqsing}{c_{\varphi q}^{(1)}}
\newcommand{\cfqtrip}{c_{\varphi q}^{(3)}}

\newcommand{\Ofu}{\mcO_{\varphi u}}
\newcommand{\Ofd}{\mcO_{\varphi d}}
\newcommand{\Ofqsing}{\mcO_{\varphi q}^{(1)}}
\newcommand{\Ofqtrip}{\mcO_{\varphi q}^{(3)}}

\newcommand{\bc}{\begin{center}}
\newcommand{\ec}{\end{center}}
\newcommand{\ba}{\begin{array}}
\newcommand{\ea}{\end{array}}

\newcommand{\pth}{p_{T}^{h}}

\def\Re{{\rm Re\,}}

\usepackage[normalem]{ulem}
\definecolor{ao}{rgb}{0.0, 0.5, 0.0}
\definecolor{awesome}{rgb}{1.0, 0.13, 0.32}

\title{Precision from the diphoton Zh channel at FCC-hh}

\author[a]{Fady~Bishara,}
\author[b]{Stefania~De~Curtis,}
\author[b]{Luigi~Delle~Rose,}
\author[a,c]{Philipp~Englert,}
\author[a,c]{Christophe~Grojean,}
\author[a,d,e]{Marc~Montull,}
\author[b]{Giuliano~Panico}
\author[a, c]{and Alejo~N.~Rossia}
\affiliation[a]{Deutsches Elektronen-Synchrotron (DESY), D-22607 Hamburg, Germany}
\affiliation[b]{Universit\`{a} di Firenze and INFN Firenze, Via Sansone 1, 50019 Sesto Fiorentino, Florence, Italy}
\affiliation[c]{Institut f{\"u}r Physik, Humboldt-Universit{\"a}t zu Berlin, D-12489 Berlin, Germany} 
\affiliation[d]{Paul  Scherrer  Institut,  CH-5232  Villigen  PSI,  Switzerland}
\affiliation[e]{Physik-Institut,  Universit{\"a}t  Z{\"u}rich,  Winterthurerstrasse  190,  CH-8057  Z{\"u}rich,  Switzerland}

\emailAdd{fady.bishara@desy.de}
\emailAdd{decurtis@fi.infn.it}
\emailAdd{luigi.dellerose@fi.infn.it}
\emailAdd{philipp.englert@desy.de}
\emailAdd{christophe.grojean@desy.de}
\emailAdd{marc.montull@psi.ch}
\emailAdd{giuliano.panico@unifi.it}
\emailAdd{alejo.rossia@desy.de}

\date{\today}
\abstract{
The future 100 TeV FCC-hh hadron collider will give access to rare but clean final states which are out of reach of the HL-LHC.
One such process is the $Zh$ production channel in the $(\nu\bar{\nu} / \ell^{+}\ell^{-})\gamma\gamma$ final states.
We study the sensitivity of this channel to 
the $\Ohq$, $\Ohqt$, $\Ohu$, and $\Ohd$ SMEFT operators,
which parametrize deviations of the $W$ and $Z$ couplings to quarks, or, equivalently, anomalous trilinear gauge couplings (aTGC).
While our analysis shows that good sensitivity
is only achievable for $\Ofqtrip$, we demonstrate that binning in the $Zh$ rapidity has the potential to improve the reach on $\Ofqsing$. Our estimated bounds are one order of magnitude better than projections at HL-LHC and is better than
global fits at future lepton colliders.
The sensitivity to $\Ofqtrip$ is competitive with other channels that could probe the same operator at FCC-hh.
Therefore, combining the different diboson channels sizeably improves the bound on $\Ohqt$, reaching a precision of $|\delta g_{1z}| \lesssim 2 \times 10^{-4}$ on the deviations in the $ZWW$ interactions.
}

\keywords{}

\begin{document}
\begin{flushright}
DESY 20-200\\
HU-EP-20/35\\
ZU-TH-43/20\\
PSI-PR-20-20
\end{flushright}

\maketitle
\flushbottom

\section{Introduction}
\label{sec.intro}

In the coming decades, future high-energy colliders will provide the next step in the study of the Standard Model (SM) and in the exploration of new physics. An accurate preliminary assessment of their physics potential is, therefore, necessary to select the most promising machines and set up a tailored and effective experimental strategy.

Among the various proposals for future accelerators, high-energy hadron colliders play a dominant role due to the broad range of new-physics probes they allow for. Although their primary target is indubitably the direct search for new massive particles, they also offer a complementary way to look for new physics through indirect probes. The latter strategy already proved successful at the LHC, where indirect electroweak (EW) probes were able in some cases to surpass the accuracy achieved at LEP~\cite{Farina:2016rws,Franceschini:2017xkh}. 

The success of the indirect search strategy is based on the fact that new-physics corrections, below the threshold for direct production, tend to grow with the energy of the process. Hadron colliders have the advantage of probing a large range of energies, thus allowing us to test the high-energy tails of kinematic distributions, where new physics is more easily accessible. The exploitation of selected ``clean'' processes with small statistical and systematic uncertainty is key for this kind of precision studies, especially when they target the EW dynamics~\cite{Farina:2016rws,Falkowski:2015jaa,Azatov:2017kzw,Franceschini:2017xkh,Bellazzini:2018paj,Grojean:2018dqj,Liu:2019vid,Azatov:2019xxn,Banerjee:2019pks,Chiu:2019ksm,Banerjee:2019twi,Freitas:2019hbk,Ricci:2020xre,Henning:2018kys,Bishara:2020vix,Chen:2020mev}.

One of the main limiting factors in carrying on the EW precision program at the LHC is the limited amount of events in many clean channels, which forces one to focus only on final states with high cross-section. 
Some of the best probes for new physics come from processes involving gauge bosons and/or the Higgs, most notably the di-boson production channels~\cite{Falkowski:2015jaa,Butter:2016cvz,Azatov:2017kzw,Franceschini:2017xkh,Liu:2019vid,Bellazzini:2018paj,Biekotter:2018rhp,Banerjee:2018bio,Grojean:2018dqj, Baglio:2018bkm,Azatov:2019xxn, Brehmer:2019gmn, Banerjee:2019pks,Banerjee:2019twi,Chiu:2019ksm,Baglio:2019uty, Bishara:2020vix,Chen:2020mev,  Baglio:2020oqu}. These processes are sensitive to many operators that describe the couplings of the SM gauge bosons and the high-energy dynamics of the Higgs. At the LHC only channels with large decay branching ratios are accessible.
For the Higgs, in particular, this means that the analysis is basically limited to the $h \to b\bar b$ decay mode, with the necessity to cope with large backgrounds and a relatively complicated final state.
For instance reconstructing an energetic Higgs in the $b\bar b$ channel requires the use of sophisticated boosted-Higgs analysis techniques.

Future high-energy hadron colliders offer a big advantage from this point of view, since they 
provide significantly larger cross-sections and much higher integrated luminosity.
This allows one to broaden the set of decay channels suitable for high-energy precision probes.
An intriguing possibility is to consider the associated production of a Higgs with a gauge boson, in which the Higgs decays in an easily reconstructable final state. For instance the $h \to \gamma\gamma$ decay mode, especially if combined with a leptonic decay of the EW boson,
can offer a very clean signature with virtually no background and small systematic uncertainties.

In a previous paper by a subset of the authors, Ref.~\cite{Bishara:2020vix}, we already started an analysis of this kind of processes focusing on $Wh$ production at a future 100 TeV proton-proton collider (FCC-hh).
We considered the final state in which the Higgs decays into a pair of photons, while the $W$ boson decays to light leptons (electrons and muons).
We found that good sensitivity to new-physics effects that modify the coupling of the $W$-boson to quarks can be achieved.
Interestingly, the expected accuracy can surpass the one achievable at the end of the LHC program by one order of magnitude and is competitive with EW precision measurements at future high-energy lepton colliders~\cite{Bishara:2020vix} (for instance FCC-ee).

In this paper, we continue the study of the associated Higgs production channels, turning our attention to the $Zh$ production process. As in Ref.~\cite{Bishara:2020vix}, we focus on the channel in which the Higgs decays into a pair of photons.
For the $Z$ boson, we consider two possible decay channels, namely the one into a pair of light charged leptons and the one into neutrinos. As we will show, both channels offer very distinctive signatures that can be used to make the analysis almost background free.

To parametrize the new-physics effects, we adopt the Effective Field Theory (EFT) approach, assuming that the beyond-the-SM (BSM) dynamics has an intrinsic scale sufficiently higher than the range of energy spanned by the $Zh$ events. As we will show, in the case of FCC-hh, this amounts to the requirement that the EFT cut-off is $\gtrsim2\;\textrm{TeV}$ (see Fig.~\ref{fig:fit_cphiq}).

With respect to $Wh$ production, the $Zh$ process is sensitive to a wider set of energy-growing new-physics effects. In fact it can be used to test the same operator, $\Ofqtrip$, that gives the leading corrections in $Wh$. Moreover it gives access to three additional operators, $\Ofqsing$, $\Ofu$ and $\Ofd$, that encode deviations in the $Z$-boson coupling to left-handed and right-handed quarks.
All these operators give rise to corrections that grow quadratically with the center-of-mass energy of the process, and are thus more visible in the high-energy tail of the kinematic distribution.

\medskip

This paper is structured as follows. In Section~\ref{sec:EFT}, we discuss the properties of the $Zh$ production channel and the leading new-physics effects it is sensitive to.
In Section~\ref{sec:signal_background}, we discuss the details of our analysis. In particular, we present the features of the signal and background processes and the cuts we exploited to single out the new-physics effects from the backgrounds.
In Section~\ref{sec.results}, the results of our analysis are collected,
whereas the conclusions of our work and possible future research directions are discussed in Section~\ref{sec:conclusions}.
We collect in Appendices~\ref{app:HelAmps} and \ref{app:mc_evt_gen} some explicit formulae and additional details about the analysis that were not included in the main text.

\section{Theoretical background}
\label{sec:EFT}
\subsection{Leading new-physics contributions}
\label{sec:interference}
We parametrize new-physics effects via the 
SMEFT formalism, 
focusing on the leading
BSM effects due to dimension-6 operators. 
Only four operators lead to contributions to the $pp\to Zh$ amplitudes that grow quadratically with the partonic center-of-mass energy~\cite{Franceschini:2017xkh}. In the Warsaw basis~\cite{Grzadkowski:2010es}, they are given by
\begin{eqnarray}
{\cal O}_{\varphi q}^{(1)} &=&\left(\overline{Q}_{L}  \gamma^{\mu} Q_{L}\right)\left(i H^{\dagger}  \overset{\leftrightarrow}{D}_{\mu} H\right)\,,\label{eq:Ofq1} \\
{\cal O}_{\varphi q}^{(3)} &=&\left(\overline{Q}_{L} \sigma^{a} \gamma^{\mu} Q_{L}\right)\left(i H^{\dagger} \sigma^{a} \overset{\leftrightarrow}{D}_{\mu} H\right)\,, \\
\Ohu &=&\left(\overline{u}_{R}\gamma^{\mu} u_R\right)\left(i H^{\dagger} \overset{\leftrightarrow}{D}_{\mu} H\right)\,, \\
\Ohd &=&\left(\overline{d}_{R}\gamma^{\mu} d_R\right)\left(i H^{\dagger} \overset{\leftrightarrow}{D}_{\mu} H\right)\,,\label{eq:Ofd}
\end{eqnarray}
where $\overset{\leftrightarrow}{D}_{\mu}\; =\overset{\rightarrow}{D}_{\mu}- (\overset{\leftarrow}{D}_{\mu})^\dagger$ 
and $\sigma^a$ are the Pauli matrices.
We define the Wilson coefficients of the above operators to be dimensionless so that the effective Lagrangian can be written as
\begin{equation}
    \lag_\textsc{eff} =\sum_a\frac{c_a}{\Lambda^2}\mcO_a\,,
\end{equation}
where $\Lambda$ is the EFT scale and the index $a$ runs over the operators in Eqs.~\eqref{eq:Ofq1}--\eqref{eq:Ofd}. To report all the numerical results we will fix the EFT scale to the conventional value $\Lambda = 1\;\textrm{TeV}$.

While, in principle, each operator also carries a generation index, we focus on the flavor-universal scenario in which the operators couple diagonally and with the same strength to all quarks. This assumption is physically motivated by the reduction of flavor constraints, in particular the stringent ones coming from flavor-changing processes. 
As we will argue later on, the bounds from our analysis mostly come from the couplings to first-generation quarks. Therefore, our results remain valid to good approximation if the operators couple only (or mainly) to the up and down quarks.

We also neglect any modification to the Higgs and $Z$-boson branching ratios, specifically $h\rightarrow \gamma\gamma$, $Z\rightarrow \ell^+ \ell^-$, $Z\rightarrow \tau^{+} \tau^{-}$, and $Z\rightarrow \nu \bar{\nu} $. This approximation is justified by the fact that, by the end of the FCC-hh program, these branching ratios are expected to be known with a precision comparable to (or below) the luminosity uncertainty.
The branching ratios of the $Z$ boson are currently measured with per mille precision~\cite{10.1093/ptep/ptaa104}.\footnote{The operators studied here modify the partial width of the $Z$ boson to quarks. The modification of the width is of order $1\%$
if we saturate the $\chq$, $\chu$ or $\chd$ bounds, and smaller than $0.3\%$ for $\chqt$.
This modification can be reconciled with the SM branching-ratio predictions via (structural) cancellations with other dimension-6 operators that contribute to the same decays. These additional operators do not induce energy-growing corrections in the $Zh$ process, but at most an overall rescaling. Since the sensitivity of our analysis comes from the shape of the differential
distribution, rather than its normalization, the bounds we derive are not significantly affected by their exclusion.
} 
On the other hand, the bound on the deviations in the effective $h\gamma\gamma$ coupling is expected to be at most $2\%$ by the end of HL-LHC and below $0.3\%$ by the end of FCC-ee+FCC-hh~\cite{deBlas:2019rxi}.
We also note that, due to the energy enhancement, the sensitivity to the effective operators~\eqref{eq:Ofq1}--\eqref{eq:Ofd} mainly comes from the differential distribution in the process energy. The overall cross-section, which is dominated by the low-energy phase-space region, is only weakly affected by new-physics contributions and plays a marginal role in the fit.

We are mainly interested in studying small deviations from the SM, i.e., the case where the SM contribution to the amplitude dominates over the BSM one.
In such a case, the main deviations from the SM predictions are expected to come from the interference between the SM and the new-physics contributions, since the square of the BSM amplitude is of higher order in the EFT expansion and therefore more suppressed.
Optimizing the sensitivity to the interference terms is therefore a crucial step in the analysis.
This generic picture might fail in specific situations, however.
As we will see in the following, in some cases the interference terms might be small or altogether absent due to structural or accidental cancellations.
When this happens, new-physics effects come dominantly from squared BSM amplitudes which leads to two additional complications.
First, new-physics effects become measurable only for relatively large values of the Wilson coefficients. And, second, additional contributions arising from dimension-8 operators, which we do not consider in our analysis, could play an important role and make the bounds more model dependent~\cite{Contino:2016jqw}.

\subsection{Structure of the interference terms}\label{sec:interf_terms}

We start by discussing the high-energy behavior of the leading-order (LO) $pp \to Zh$ helicity amplitudes. 
The scaling with the center-of-mass energy of the process, $\hat s$, is shown in Table~\ref{tab:Operator_growth}. The full expressions for the LO helicity amplitudes are reported in Appendix~\ref{app:amps_wh}.

\begin{table}[t]
\begin{center}
\begin{tabular}{ c | c@{\hspace{1.5em}}c }
  \toprule
  $Z$ polarization & SM & ${\cal O}_{\varphi q}^{(3)}, {\cal O}_{\varphi q}^{(1)}, {\cal O}_{\varphi u}, {\cal O}_{\varphi d}$\\
  \midrule
  $\lambda=0$ & $1$  & $\dfrac{\hat{s}}{\Lambda^2}$\\[4mm]
  $\lambda=\pm$ & $\dfrac{M_Z}{\sqrt{\hat{s}}}$ & $\dfrac{\sqrt{\hat{s}}\,M_Z}{\Lambda^2}$\\
  \bottomrule
\end{tabular}
\end{center}
\caption{High-energy behavior of the tree-level SM and BSM helicity amplitudes for the process $q \bar q \rightarrow Zh$.}
\label{tab:Operator_growth}
\end{table}

For large partonic center-of-mass energy, $\sqrt{\hat{s}} \gg M_Z$, the SM amplitude with a longitudinally polarized $Z$ boson is constant and dominates over the transversely polarized ones, which are suppressed by  a relative factor of $M_Z/\sqrt{\hat{s}}$.

On the other hand, all the contributions corresponding to the dimension-6 operators~\eqref{eq:Ofq1}--\eqref{eq:Ofd} exhibit the same behavior.
The amplitudes with a longitudinally polarized $Z$ boson grow as $\hat{s}/\Lambda^2$, i.e., with the square of the center-of-mass energy, while, for a transversely polarized $Z$, the growth is only linear, of order $\sqrt{\hat{s}} M_Z/\Lambda^2$.

A similar behavior characterizes the closely related $Wh$ process~\cite{Bishara:2020vix}.
By contrast, in the $WZ$ case, the transverse opposite-polarization channels are unsuppressed in the SM and constitute an important background for the BSM signal~\cite{Franceschini:2017xkh}.

If one performs a fully inclusive analysis in the $Z$ decay products, only amplitudes with the same $Z$ polarization can interfere.
In this case, the behavior of the squared SM amplitude and of the leading interference terms is\footnote{The full expressions for the squared SM amplitude and the interference terms are given in Appendix~\ref{app:amps_squared}.}
\begin{equation}
\begin{split}
   \left| \mcM_\textsc{sm} \right|^2 &\sim \sin^2 \theta \,,\\
   \Re\mcM_{SM} \, \mcM_{\textsc{bsm}}^* &\sim \frac{\hat s}{\Lambda^2} \sin^2 \theta\,,
\end{split}\label{eq:main_amp_sq}
\end{equation}
where $\mcM_\textsc{bsm}\in\{\mcM_{\varphi q}^{(3)},\mcM_{\varphi q}^{(1)}, \mcM_{\varphi u}, \mcM_{\varphi d}\}$ and $\theta$ is the scattering angle of the $Z$ boson.
That is, all the leading BSM contributions can directly interfere with the leading SM amplitude.
Thus, the main BSM effects are expected to be captured by exploiting the transverse-momentum distribution which is closely related to the $\hat s$ distribution.

However, stark differences in the size of the interference terms are present even though the behavior of the BSM amplitudes is the same for all effective operators~\eqref{eq:Ofq1}--\eqref{eq:Ofd}.
This can be seen by direct inspection of the analytic expressions (see Appendix~\ref{app:HelAmps}) and numerical analyses (see Tables~\ref{tab:sigma_full} and~\ref{tab:sigma_full_lep}).
The only operator that leads to a sizeable interference is $\Ofqtrip$, while all others suffer important suppressions.

For the right-handed operators, $\Ofu$ and $\Ofd$, the interference terms are suppressed since they are proportional to the coupling of the $Z$ boson to right-handed quarks, which is significantly smaller than the one to left-handed quarks.
BSM effects linear in the Wilson coefficients are therefore suppressed with respect to the SM contributions and to the quadratic BSM terms, degrading the sensitivity.
In fact we find that, for values of the Wilson coefficients of order of the expected bounds ($\cfu \sim \cfd \sim 2 \times 10^{-2}$), linear and quadratic BSM corrections are roughly of the same order.

For the weak isospin singlet operator, $\Ofqsing$, a (partially accidental) cancellation between the up-type and down-type quark contributions is present. This cancellation arises because the SM-BSM interference term is linear in the SM coupling of the quarks to the $Z$. Such coupling is proportional to $T_3 - s_{\textsc w}^2 Q$, where $T_3$ is the weak isospin charge, $Q$ the electric charge and $s_{\textsc w}$ is the sine of the weak mixing angle, and have opposite sign for up-type and down-type quarks.
On the other hand, the leading BSM amplitudes (see Eq.~\eqref{eq:amp_Oq1}) coming from $\Ofqsing$ have the same sign for all quarks, leading to opposite-sign interference contributions for up-type and down-type quarks.\footnote{The $\Ofqtrip$ amplitudes with up-type and down-type quarks have opposite sign (see Eq.~\eqref{eq:amp_Oq3}), which compensates the sign change in the SM coupling.}
The suppression is further exacerbated by the relative weight of the valence-quark content of the proton, which makes the cancellation stronger. 
As a result, the sensitivity to $\Ofqsing$ is quite degraded and dominated by the terms that are quadratic in the Wilson coefficients.

\begin{figure}
	\centering
	\includegraphics[scale=1]{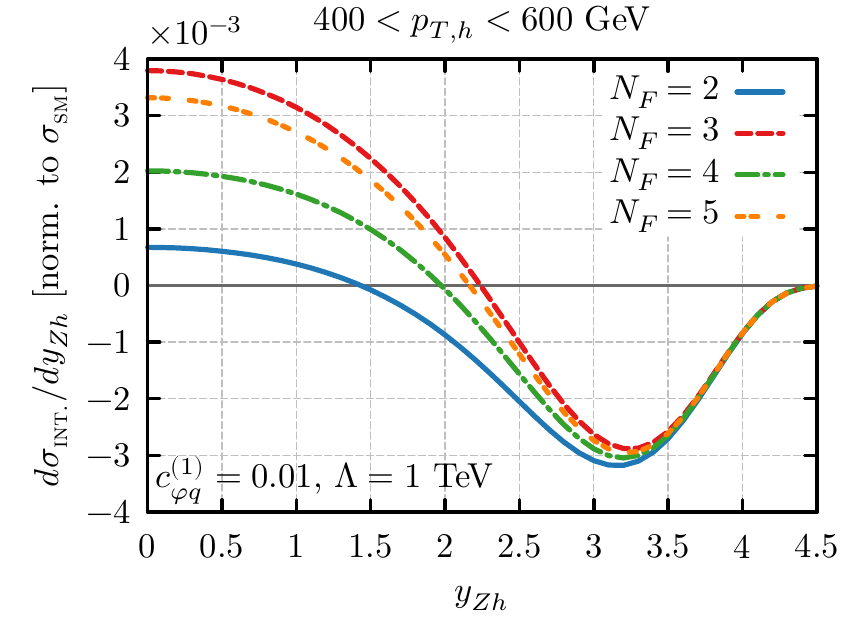}
	\caption{The differential distribution of the interference term, $\propto \cfqsing$, w.r.t. the rapidity of the $Zh$ system, $y_{Zh}$. The value of $N_F$ indicates the number of initial-state flavors summed over (in PDG order).
	The distribution is integrated over the $p_{T,h}$ range of $[400,600]$ GeV and $\sqrt{\hat{s}}$ over the appropriate range.
	The normalization factor is the SM only distribution further integrated over $y_{Zh}$.}
	\label{fig:y_binning}
\end{figure}

A possible way to partially enhance the interference terms is to consider the differential distribution in the rapidity of the $Zh$ system (or a correlated quantity, such as the Higgs rapidity). The rapidity distribution coming from up-type and down-type initiated processes is slightly different. In particular the $u \bar u$ parton luminosity is peaked at larger rapidity than the $d \bar d$ one (see for instance Fig.~3 of Ref.~\cite{Panico:2016ary}). Separating the small rapidity and large rapidity regions, the cancellation can be partially undone.
This is shown in Fig.~\ref{fig:y_binning} where the differential cross-section with respect to the $Zh$ rapidity (normalized to the total SM cross-section) is plotted for different flavor assumptions. Note that only the interference term, which is linear in the Wilson coefficient, is shown.
The solid, dash-dotted, double-dashed, and dashed curves correspond to the sum over the lightest 2, 3, 4, and 5 quarks respectively.
This term changes sign around $y_{Zh}\sim 2$ because the relative contribution between the $u$- and $d$-quark initial states depends on $y_{Zh}$ as discussed above.
The choice of $N_F$ depends on the flavor assumption imposed on the Wilson coefficient $\cfqsing$.
The solid, $N_F=2$, curve sums over first-generation quarks which comprise the valence quarks.
For any $N_F>2$, each down-type quark contributes positively in the central region ($|y_{Zh}|<2$) and each up-type quark contributes negatively. The size of the contribution progressively decreases the heavier the flavor. 
The resulting differences in the distributions could, in principle, be exploited to disentangle the flavor assumption given a clean decay channel with large statistics to sufficiently suppress the quadratic terms. 

As we will discuss in Section~\ref{sec.results}, the impact of a rapidity binning on the analysis is small, mostly because the statistics in each bin is limited. Taking into account the rapidity distribution could however be more relevant for different final states with larger cross-section. For instance the $Zh$ production process with the Higgs decaying into $b$ quarks. We leave this analysis for future studies.

To conclude the discussion, we mention that interference between different $Z$-boson polarization channels can be recovered by considering differential distributions that also include the $Z$ decay angles~\cite{Panico:2017frx}. These effects, however, do not grow with energy and are small with respect to the main interference term which does.
Taking into account the decay product distributions is also not useful to remove the suppression of the linear interference terms in $\cfqsing$, $\cfu$ and $\cfd$.
In fact all the interference amplitudes are controlled by the same combination of couplings and the same suppression patterns characterize all of them.
We conclude that a sophisticated analysis taking into account the kinematics of the Z-boson decays can be avoided without a significant loss of sensitivity.

\subsection{Additional contributions to the signal}\label{sec:additional_signal}

So far we only considered the LO contributions to the $Zh$ production channel. Additional contributions, with potentially different features, arise at the 1-loop level.
In particular, the $gg \to Zh$ channel can show a very different dependence on the effective operators~\eqref{eq:Ofq1}--\eqref{eq:Ofd}.
Although new-physics contributions do not grow with the energy in this channel~\cite{Banerjee:2018bio}, a sizeable dependence on the effective operators might still be present.
As shown in Ref.~\cite{Englert:2013vua}, the SM contribution to the $gg \to Zh$ channel is suppressed (especially at high energy) by a cancellation between the contributions from box-type and triangle-type diagrams.
The presence of new-physics contributions can partially remove the cancellation, thus leading to sizeable deviations in the $\hat s$ distribution which grow at high energy.
In fact we found that the corrections induced by $\Ohq$ for a value of the Wilson coefficient $\chq = 1.5 \times 10^{-2}$  (close to the bound we derive from our analysis) range from $\sim 4\%$ in the low-energy bins up to $\sim 60\%$ in the high-energy tail.

Extracting information from the $gg \to Zh$ channel can, however, be tricky for several reasons. First of all, its cross-section is much smaller than the leading $q\bar q \to Zh$ one and, in the high-energy bins, where the corrections are stronger, it contributes to less than one event at the end of the FCC-hh operation (see Fig.~\ref{fig:pTh}). Special cuts would therefore be needed to enhance its visibility.
Second, the $gg \to Zh$ process strongly depends on additional effective operators, whose determination could be not sufficiently precise to remove their effects. In particular, there is a strong dependence on deviations in the top Yukawa coupling.
An uncertainty on its determination at the $\sim 1\%$ level could easily overshadow the effects of the effective operators we are considering in this paper.

In the light of these difficulties, in our analysis, we will not exploit the new-physics dependence of the $gg \to Zh$ channel, ignoring its dependence on the Wilson coefficients and treating it as a background.

It is worth mentioning here another channel that can contribute to the signal. We characterize the final state with the $Z$ decaying into neutrinos, by requiring a pair of photons and missing transverse momentum. A non-negligible fraction of events with this signature comes from the $Wh$ production channel with a leptonically-decaying $W$ boson, in which the lepton is not detected. This happens in a significant fraction of the events in which the $W$ decays into taus, while it is much rarer for decays into electrons and muons. The $Wh$ channel depends only on the $\Ofqtrip$ operator~\cite{Franceschini:2017xkh,Bishara:2020vix}
with corrections that grow with the energy, thus enhancing the sensitivity to this operator.

\subsection{Power counting considerations}\label{sec:power_counting}

We conclude this section with a short discussion about the expected size of
the new-physics effects in generic BSM theories.
All four operators considered here can be generated at tree-level via fermionic or bosonic weak isospin singlet and triplet states~\cite{deBlas:2017xtg}. Therefore, from a power-counting point of view, no particular hierarchy exists between their Wilson coefficients. According to the SILH power counting~\cite{Giudice:2007fh}, we find
\begin{equation}
\chqt \sim \chq \sim \chu \sim \chd \sim g^2\,,
\end{equation}
where $g$ is the SM EW gauge coupling. This estimate is valid in theories in which new physics is weakly coupled or is not directly coupled to the SM fields.

If new physics is strongly coupled to the SM, the estimate becomes
\begin{equation}
\chqt \sim \chq \sim \chu \sim \chd \sim g_*^2\,,
\end{equation}
where $g_*$ is the typical size of the new-physics coupling.


\section{Event generation and analysis}\label{sec:signal_background}

We simulated signal and background events with \textsc{MadGraph5\_aMC@NLO} v.2.7.3~\cite{Alwall:2014hca} using the \texttt{NNPDF23} parton distribution functions~\cite{Ball:2013hta}. Parton shower and Higgs decay into two photons were modeled using \textsc{Pythia8.24}~\cite{Sjostrand:2014zea}, while detector effects were estimated via \textsc{Delphes} v.3.4.1~\cite{deFavereau:2013fsa,Selvaggi:2014mya,Mertens:2015kba,Cacciari:2011ma,Cacciari:2005hq,Cacciari:2008gp} with its FCC-hh card.
All the simulations were done using the \texttt{SMEFTatNLO}~\cite{Degrande:2020evl} \texttt{UFO} model~\cite{Degrande:2011ua},
which implements the effective operators we consider in our analysis. 
Further details about the Monte Carlo generation can be found in Appendix~\ref{app:mc_evt_gen}.

The signal process is $p p \to Zh$, with the Higgs decaying into a photon pair and the $Z$ decaying either into a pair of charged leptons, $Z \to \ell^+ \ell^-$ (with $\ell = e, \mu$), or into neutrinos, $Z \to \nu \bar \nu$.
The channel in which the $Z$ boson decays hadronically suffers from a much larger background and requires a more sophisticated analysis strategy that we leave for future work.

\subsubsection*{$Z\rightarrow \nu\bar{\nu}$ channel}
\label{sec:Relev_proc_nunu}

The main process contributing to this channel is $q\bar{q}\rightarrow Z\left(\rightarrow \nu\bar{\nu}\right)h\left(\rightarrow \gamma\gamma\right)$. The signature is given by a pair of photons and sizeable missing transverse momentum.
As we mentioned in Section~\ref{sec:additional_signal}, the additional process $q\bar{q}\rightarrow W\left(\rightarrow \ell\nu\right)h\left(\rightarrow \gamma\gamma\right)$ where the charged lepton is not reconstructed contributes to the same final state and must be considered as part of the signal, since it depends on the $\Ofqtrip$ operator. This has been found to be relatively unlikely for a $W$ to decay into electrons and muons, but happens in a non-negligible fraction of events with $W \to \tau \nu_\tau$ (see Section~\ref{sec:cut_efficiency}).
We simulated these channels at LO in QCD and we took into account NLO QCD and EW corrections through k-factors.
NLO corrections turn out to be sizeable, but with a weak dependence on new-physics.
It is thus a good approximation to rescale the cross-section in each bin by the SM NLO k-factors which are given in Table~\ref{tab:k_factors}.
We also take into account the $gg \to Z(\to \nu\bar{\nu})h(\to \gamma\gamma)$ production channel at LO but neglect new-physics contributions as discussed above.

The main backgrounds for this channel come from the processes $Z\left(\rightarrow \nu\bar{\nu}\right)\gamma\gamma$ and $W\left(\rightarrow \ell\nu\right)\gamma\gamma$, where the charged lepton is missed and the photon pair is non-resonantly produced.
Additional contributions from events in which one or two jets fake a photon are negligible if we assume a $P_{j \to \gamma} = 10^{-3}$ jet-to-photon fake rate~\cite{Contino:2016spe,Abada:2019lih} (see also discussion in Appendix B of Ref.~\cite{Bishara:2020vix}). Consequently, we do not consider processes with less than $2$ photons as part of the background. This is valid for both $Z$ decay channels.
We also checked that the $j\gamma\gamma$ process in which missing transverse momentum comes from showering and detector effects, constitutes a huge background at low energy, but becomes negligible when a $200\;\rm{GeV}$ cut on the missing transverse momentum is imposed. 

The backgrounds were generated at NLO in QCD. This is particularly important for the 
$W\left(\rightarrow \ell\nu\right)\gamma\gamma$ and $Z\left(\rightarrow \nu\bar{\nu}\right)\gamma\gamma$ channels. In the former, the cross-section is fully dominated by the NLO contribution with a real emission (which is a factor $\sim 17$ larger than the LO contribution).
Similarly, the NLO contributions for the $Z\left(\rightarrow \nu\bar{\nu}\right)\gamma\gamma$ channel are a factor $\sim 2$ larger than the LO ones.
The enhancement is partly due to the fact that at NLO an additional channel with an initial gluon instead of a sea quark opens up.
As we will see in Section~\ref{sec:cut_efficiency},
the presence of a large component of events with an additional jet significantly modifies the kinematic configuration of the events, providing efficient handles to distinguish the background processes from the signal.

\subsubsection*{$Z\rightarrow \ell^{+}\ell^{-}$ channel}
\label{sec:Relev_proc_lele}

The main signal process in this channel is $q\bar{q}\rightarrow Z\left(\rightarrow \ell^{+}\ell^{-}\right)h\left(\rightarrow \gamma\gamma\right)$, with $\ell = e, \mu$. 
The main background process is $Z(\rightarrow \ell^{+}\ell^{-})\gamma\gamma$, with a small additional contribution
from processes in which the lepton pair arises from the splitting of a virtual photon.
This contribution can be suppressed by restricting the di-lepton invariant mass to be around the $Z$-pole.
As for the neutrino channel, we generated the signal at LO QCD, applying suitable k-factors to take into account NLO QCD and EW effects (see Table~\ref{tab:k_factors}), while the background was simulated at NLO in QCD. Clearly, the NLO corrections follow a pattern completely analogous to the one we discussed in the neutrino channel.
We also took into account at LO the $gg \to Z(\to \ell^+ \ell^-)h(\to \gamma\gamma)$ channel, considering it as a constant background. 

Finally, we mention that the decay of the $Z$ boson into a pair of taus, when both taus decay fully leptonically, can also contribute to the $\ell^{+}\ell^{-}$ channel. We estimate that this process would increase the signal and background cross-sections in a similar way, leading to at most a $5\%$ increase. The impact of such change on our results would be negligible, therefore we do not include this channel in our simulations.


\subsection{Selection cuts}
\label{sec:cuts}

In order to reconstruct the signal we require the presence of at least two photons with $p_{T}>50$ GeV and $|\eta|<6$. In addition, at least one pair of photons must have an invariant mass in the range $120\;\textrm{GeV}<m_{\gamma\gamma}<130\;\textrm{GeV}$. 
If more than one pair of photons fulfill that requirement, we choose the pair with the smallest angular separation, $\Delta R_{\gamma\gamma} = \sqrt{\left(\Delta\eta_{\gamma\gamma}\right)^2+\left(\Delta\phi_{\gamma\gamma}\right)^2}$.~\footnote{This selection procedure could in principle be affected by radiated photons. We however verified that this effect is negligible. We also verified that the alternative procedure of selecting the pair with the hardest photons gives almost equal results.}

In the $Z\rightarrow \nu\bar{\nu}$ channel, we veto events with muons or electrons in the kinematic region defined by $p_{T}>30$ GeV and $|\eta|<6$.
In the $Z\rightarrow \ell^+ \ell^-$ channel, we only accept events which contain $2$ leptons of opposite charge, each with $p_{T}>30$~GeV and $|\eta|<6$.
The invariant mass of the lepton pair $m_{l^{+}l^{-}}$ is required to be in the range $[81, 101]$ GeV to ensure the reconstruction of the $Z$ boson mass.

\begin{table}[t]
	\centering{
		\renewcommand{\arraystretch}{1.25}
		\begin{tabular}{ c @{\hspace{.5em}} | @{\hspace{.5em}} c   }
			\toprule
			& Selection cuts  \\\midrule
			$p_{T,\min}^{\ell}$ [GeV] &  30 \\
			$p_{T, \min}^{\gamma}$  [GeV] & 50 \\
			$m_{\gamma \gamma}$ [GeV] & $[120,130]$ \\
			$m_{\ell^{+} \ell^{-}}$ [GeV] & $[81,101]$ \\
			\hline
			$p_{T,\max}^{Zh}$ [GeV] &  $\{200,600,1100, 1500, 1900\}$  \\
			\bottomrule
		\end{tabular}}
		\caption{Summary of the selection cuts. The cut $m_{\ell^{+} \ell^{-}}$ is applied only in the $Z\rightarrow\ell^+ \ell^-$ channel, the others are common to both channels. The entries for $p_{T,\max}^{Zh}$ correspond to different generation bins in $\pth$, $p_{T}^{Z}$ and $p_{T,\textrm{min}}$.
		}
		\label{tab:sel_cuts}
	\end{table}

Finally, we impose a maximum cut on the transverse momentum of the reconstructed $Zh$ system, $p_{T}^{Zh} < p_{T,\max}^{Zh}$. This cut is motivated by the fact that a large transverse momentum is typically the signal of a recoil against a hard QCD jet. Such events are more likely to come from background processes than from the signal.
A summary of the selection cuts and of the values of
$p_{T,\max}^{Zh}$ is given in Table~\ref{tab:sel_cuts}.

To conclude this discussion, let us mention that we also checked the usefulness of an upper cut on the angular separation of the photon pair, $\Delta R_{\gamma\gamma}$, and of the lepton pair, $\Delta R_{\ell^{+} \ell^{-}}$. These cuts are motivated by the fact that the di-photon and di-lepton pairs, coming from boosted objects, typically have smaller $\Delta R$ than the non-resonant background processes. We however found that, after the cuts on the invariant masses,
the $\Delta R$ distributions for signal and backgrounds were very similar, so that a cut on the angular separation does not improve the fit.

\subsection{Binning}
\label{sec:binning}

In our analysis, we consider the double-differential distribution in the center-of-mass par\-to\-nic-energy and in the rapidity of the events (see Section~\ref{sec:EFT}).
As can be seen from Eq.~\eqref{eq:main_amp_sq}, the signal events are mostly emitted in the central scattering region. We can therefore approximately trade the event energy for the transverse momentum of the $Z$ or Higgs.
Both of these can be reconstructed in the two decay channels we consider.
We choose to bin in the minimum $p_T$ of the two bosons
\begin{equation}
p_{T,\textrm{min}} = \textrm{min}\{p_T^h, p_T^Z\}\,,
\end{equation}
where, in the neutrino channel, $p_T^Z$ is identified with the missing transverse momentum.
This choice is useful to select hard events in the $Zh$ center-of-mass frame and, simultaneously, to remove the $j\gamma\gamma$ background where the missing energy comes from soft radiation.
We use five bins in $p_{T,\textrm{min}}$, whose boundaries are given by
\begin{equation}
p_{T,\textrm{min}} \in \{200, 400, 600, 800, 1000, \infty\}\;\textrm{GeV}\,.
\end{equation}

In the $Z \to \ell^+ \ell^-$ channel, we also use a simple binning in the rapidity of the $Zh$ system $y_{Zh}$, namely
\begin{equation}
|y_{Zh}| \in [0,2], [2, 6]\,,
\end{equation}
while in the $Z \to \nu \bar{\nu}$ channel, since the rapidity of the $Zh$ system cannot be determined, we bin in the rapidity of the Higgs $y_h$ instead, which is strongly correlated with $y_{Zh}$. The bin definitions depend on $p_{T,\min}$ as follows,
\begin{equation}
\left|y_h\right|\in
    \begin{cases}
    [0,2],\,[2,6] & \quad\text{for}\quad 200<p_{T,\min} <600\;\text{[GeV]}\,,\\
    [0,1.5]\,,[1.5,6] & \quad\text{for}\quad 600<p_{T,\min} <800\;\text{[GeV]}\,,\\
    [0,1]\,,[1,6] &\quad\text{for}\quad  p_{T,\min}>800\;\text{[GeV]}.
    \end{cases}
\end{equation}

\subsection{Cut efficiencies}\label{sec:cut_efficiency}

We show in Tables~\ref{tab:Cutflow_Nu} and \ref{tab:Cutflow_Le} the cutflow analysis for the different signal and backgorund processes in the $Z\rightarrow \nu \bar{\nu}$ and $Z\rightarrow \ell^{+} \ell^{-}$ channels. In order to focus on events that have a good sensitivity to new-physics, in obtaining these results we only considered high-energy events that satisfy the parton-level cut $p_T^{Z,W} > 400\;\textrm{GeV}$.
Notice that the initial phase space of the different processes does not exactly coincide, due to small differences in the generation-level cuts (see Appendix~\ref{app:mc_evt_gen} and Table~\ref{tab:gen_cuts}). The results presented in the tables are however still useful to understand the efficiency of each cut in suppressing the backgrounds.

As expected, the cut on the invariant mass of the photon pair, $m_{\gamma\gamma}$, is very efficient in reducing all backgrounds with non-resonant photon pairs, i.e., $W\gamma\gamma$ and $Z\gamma\gamma$.
Analogously, the cut on the lepton-pair invariant mass $m_{\ell^+\ell^-}$ is helpful in reducing the $\ell^+\ell^-\gamma\gamma$ background component in which the lepton pair comes
from an off-shell photon line. Such channel becomes completely negligible after the cuts, so that we did not include it in Table~\ref{tab:Cutflow_Le}, where only the $Z\gamma\gamma \to \ell^+\ell^-\gamma\gamma$ channel is listed.

\begin{table}
\begin{centering}
\renewcommand{\arraystretch}{1.25}
\begin{tabular}{c|c|c|c|c|c}
\toprule
Cuts / Efficiency & $q\bar{q}\rightarrow Zh$ & $Wh$ &$W\gamma\gamma$ & $Z\gamma\gamma$ & $gg\rightarrow Zh$\tabularnewline
\midrule
$0$ $\ell^{\pm} $ in acc. region  & $1$ & $0.30$  & $0.44$ & $1$ & $0.97$ \tabularnewline
$\geq2\,\gamma$ in acc. region  & $0.60$ & $0.19$ & $0.30$  & $0.72$ & $0.60$ \tabularnewline
$m_{\gamma\gamma} \in [120,\, 130]$\,GeV & $0.58$ & $0.17$ & $7.7 \times 10^{-3}$ & $1.3 \times 10^{-2}$ & $0.59$\tabularnewline
$p_{T,\textrm{min}}\geq 400$ GeV & $0.42$ & $0.061$ & $6.9\times10^{-4}$ & $2.9\times10^{-3}$ & $0.37$\tabularnewline
$p_{T}^{Zh}\leq p_{T,max}^{Zh}$ & $0.40$ & $0.057$  & $1.1\times10^{-4}$ & $2.8\times10^{-3}$ & $0.33$\tabularnewline
\bottomrule
\end{tabular}
\par\end{centering}
\caption{Cutflow for the processes in the $Z\rightarrow \nu\bar{\nu}$ channel. The acceptance region for charged leptons and photons is defined in the text.}
\label{tab:Cutflow_Nu}
\end{table}

\begin{table}
\begin{centering}
\renewcommand*{\arraystretch}{1.25}
\begin{tabular}{c|c|c|c}
\toprule
Cuts / Efficiency & $q\bar{q}\rightarrow Zh\rightarrow \ell^{+}\ell^{-}\gamma\gamma$ & $Z\gamma\gamma \to \ell^{+}\ell^{-}\gamma\gamma$ & $gg\rightarrow Zh\rightarrow \ell^{+}\ell^{-}\gamma\gamma$\tabularnewline
\midrule
$2$ $\ell^{\pm}$ in acc. region & $0.85$ & $0.74$ & $0.75$\tabularnewline
$\geq2\,\gamma$ in acc. region & $0.51$ & $0.54$ & $0.46$ \tabularnewline
$m_{\gamma\gamma} \in [120,\, 130]$ GeV & $0.50$ & $9.4\times10^{-3}$ & $0.45$ \tabularnewline
$m_{l^{+}l^{-}} \in [81, 101]$ GeV & $0.47$ & $8.8 \times 10^{-3} $ & $0.42$ \tabularnewline
$p_{T,\textrm{min}}\geq 400$ GeV & $0.35$ & $2.2 \times 10^{-3}$ & $0.26$ \tabularnewline
$p_{T}^{Zh}\leq p_{T,max}^{Zh}$ & $0.33$ & $2.1\times10^{-3}$ & $0.23$ \tabularnewline
\bottomrule
\end{tabular}
\par\end{centering}
\caption{Cutflow for the processes in the $Z\rightarrow \ell^{+} \ell^{-}$ channel. The acceptance region for charged leptons and photons is defined in the text.}
\label{tab:Cutflow_Le}
\end{table}

It is interesting to notice that, in the neutrino channel, the charged-lepton veto reduces the cross-section of the $Wh$ and $W \gamma \gamma$ channels by only $60 - 70\%$. This behavior, which might seem surprising at first sight, is mainly due to the $W \to \tau \nu_\tau$ decay channel, in which the $\tau$ decays hadronically.
On the other hand, when the $W$ decays to light leptons, the cut is much more efficient, with a reduction of the events of order $96\%$ for decays with an electron and $98\%$ for decays with a muon.

A large fraction of the events with hadronically decaying taus is however removed by the cuts on $p_{T, min}$ and $p_T^{Zh}$. The reason for this is twofold.
First, the jets from the $\tau$ decay carry a sizeable fraction of the $\tau$ momentum, thus creating an unbalance between the missing transverse momentum and the photon pair $p_T$.
And, second, in the $W\gamma\gamma$ channel, a vast majority of the events come from the NLO contribution with an extra hard jet. In such events a large transverse momentum $p_T^{Zh}$ is present.

The $p_{T, min}$ cut is also quite efficient in reducing the $Z\gamma\gamma$ backgrounds both in the neutrino and lepton channels.
This, again, is due to the fact that many events ($\sim 65\%$) come from the NLO contribution with an additional hard jet.

\begin{figure}
	\centering
	\includegraphics[width=0.495\linewidth]{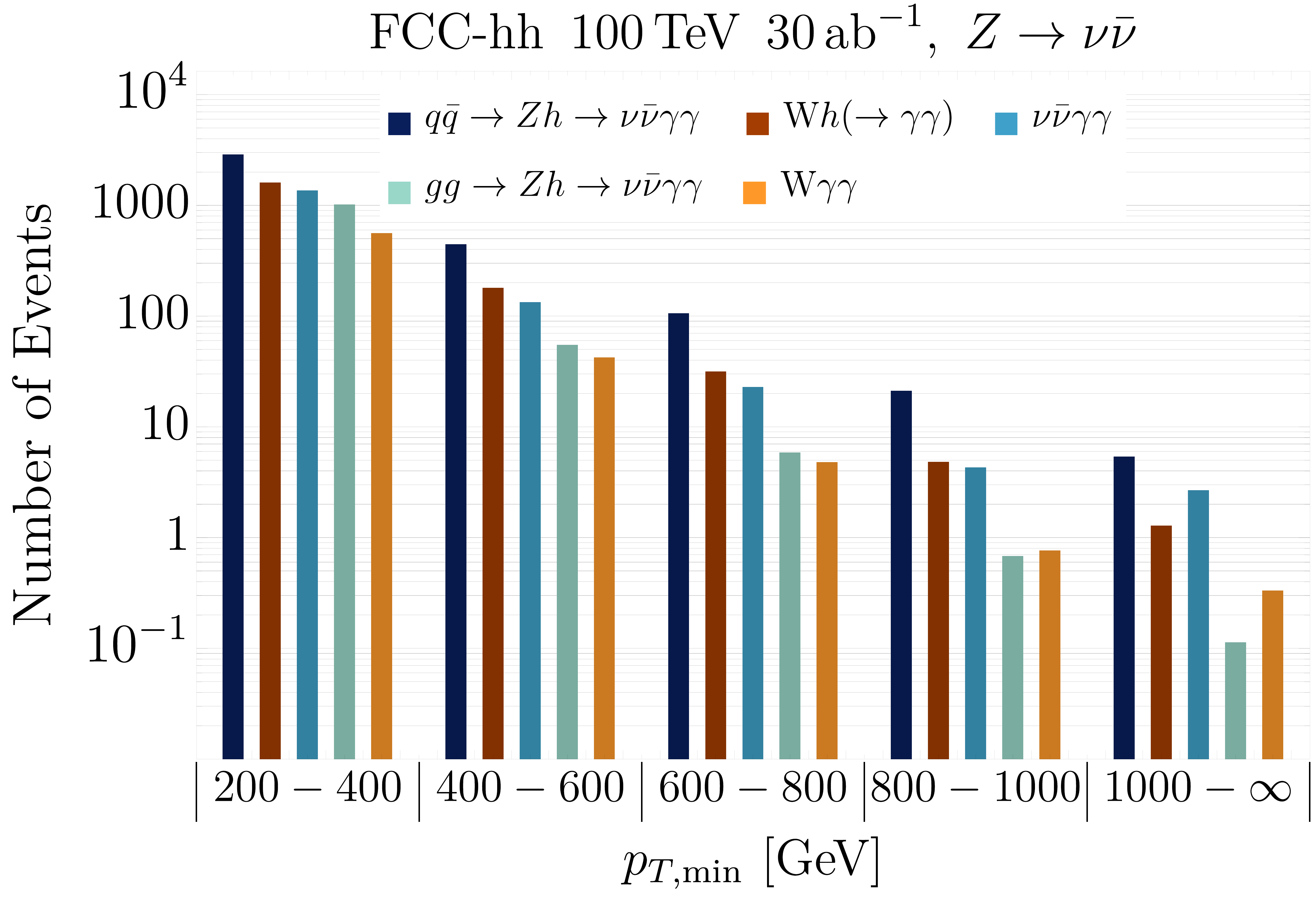}    
	\hfill
	\includegraphics[width=0.495\linewidth]{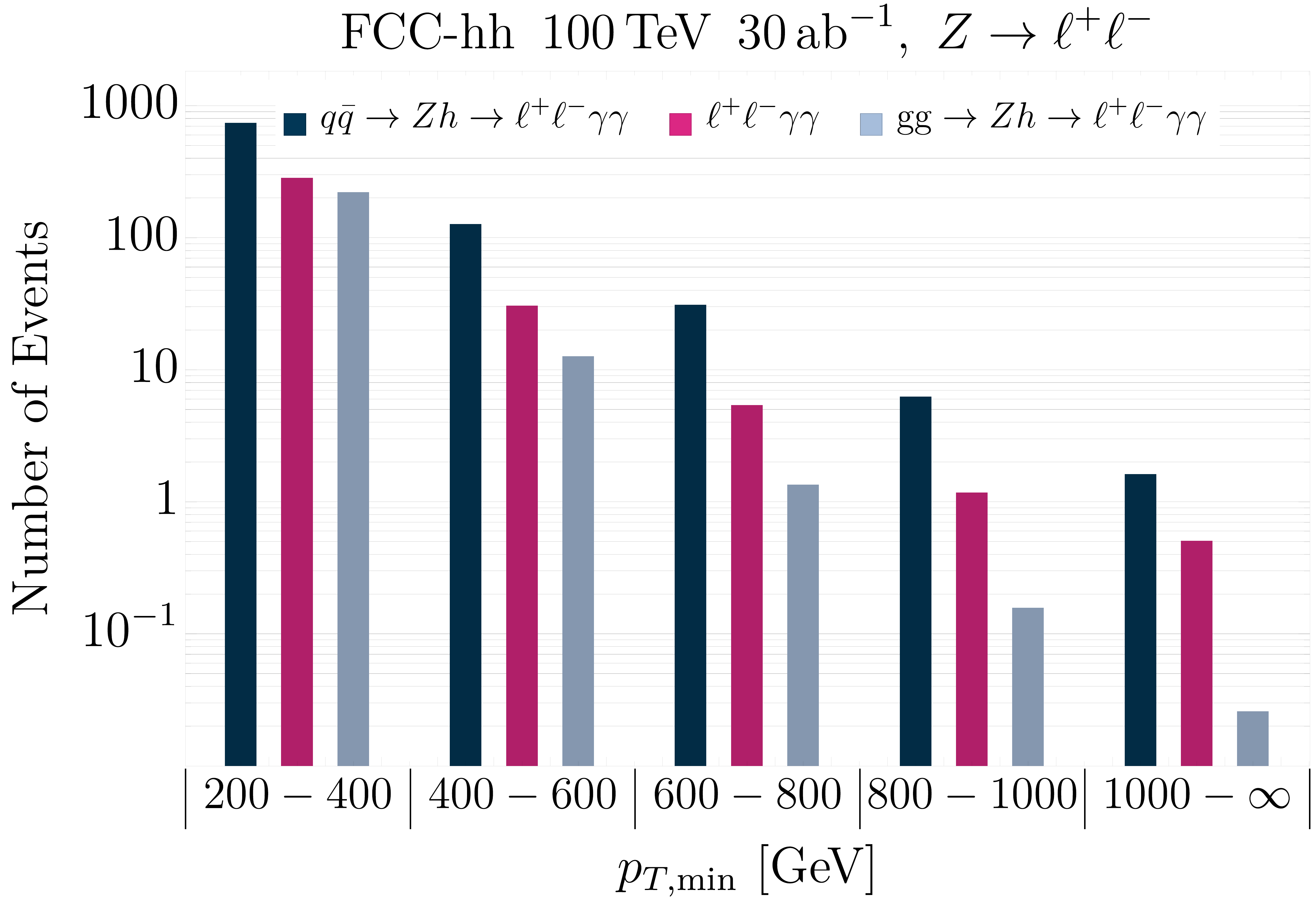}
	\caption{Number of SM events per bin after selection cuts for the signal and backgrounds at the FCC-hh assuming $\mathcal{L}=30$ ab$^{-1}$ integrated luminosity. Left (right) panel: for the decay channel $Z\rightarrow \nu\bar\nu$ ($Z\rightarrow \ell^+\ell^-$), where the bins are defined according to $p_{T,\textrm{min}}$. 
	}
	\label{fig:pTh}
\end{figure}

In Fig.~\ref{fig:pTh}, we show the number expected of SM events at the FCC-hh for the signal and background processes in each $p_{T,\textrm{min}}$ bin.
For the sake of clarity, the figure only shows the sum of the two rapidity bins.
The number of signal and background events in all bins used in the analysis are listed in Tables~\ref{tab:sigma_full} and \ref{tab:sigma_full_lep}, together with the dependence of the signal cross-section on the Wilson coefficients.

In the neutrino channel, it is remarkable that the $Wh$ channel gives a significant contribution to the signal with a number of events of the same order of magnitude as in the $Zh$ channel.
This is true in spite of the fact that the $Wh$ channel has a much lower cut efficiency with respect to $Zh$, as can be seen from Table~\ref{tab:Cutflow_Nu}.
The suppression is however partially compensated by the significantly higher initial cross-section.
Moreover the dependence on $\cfqtrip$ in the $Wh$ channel is somewhat stronger than in $Zh$, due to accidental numerical factors in the new-physics amplitude.
For this reason the $Wh$ channel can enhance, up to $\sim 50\%$, the dependence of the cross-section on $\cfqtrip$, as can be seen
by comparing the results in Tables~\ref{tab:sigma_full} and \ref{tab:sigma_full_lep}. 

The main background in the neutrino channel is $Z\gamma\gamma \to \nu\bar{\nu}\gamma\gamma$,
followed by $W\gamma\gamma$. In most of the $p_{T,min}$ bins, the backgrounds are much smaller than the signal, making the analysis almost background free. The only exceptions are the first bin, where however the dependence on new physics is small, and the last one, where the number of events is quite limited. Finally the $gg\rightarrow Zh$ channel is also quite small, so that its dependence on new physics can be safely neglected.

In the charged lepton channel, the $Z\gamma \gamma \to \ell^{+}\ell^{-}\gamma\gamma$ background is much smaller than the signal in all bins. Analogously to the neutrino channel, the $gg\rightarrow Zh$ has an almost negligible impact, especially in the higher $p_{T,min}$ bins.

Before discussing the results, we make a comment on the impact of the flavor-uni\-ver\-sa\-li\-ty assumption for the new-physics contributions. The main signal channels $Zh$ and $Wh$ are dominated by the production through initial states containing first-generation quarks.
Contributions from second-generation quarks and from the bottom are suppressed by the proton parton distribution functions, and together give $\sim 20\%$ of the whole cross-section.~\footnote{The relative contribution of each $q\bar{q}$ initial state to the total cross-section at LO in the bin $400\;\textrm{GeV}<p_{T,h}<600\;\textrm{GeV}$ is: $u(42\%),\;d(38\%),\;s(9.2\%),\;c(5.7\%),\;b(4.7\%)$.}
Thus, the analysis we perform approximately captures the case in which new physics couples only the first generation, or the case of $\textrm{U}(2)$ flavor symmetry in the first two generations.

The situation is potentially different for the $gg \to Zh$ channel. In this case new-physics corrections to the top-quark couplings can strongly affect the dominant loop diagrams. In the flavor-universal scenario, the corrections to the top couplings are related to the ones to the light generations, thus they are relatively small. We checked that, for values of the Wilson coefficients of the order of the bounds we find ($|\cfqtrip| \sim 3 \times 10^{-3}$; $|\cfqsing|, |\cfu|, |\cfd| \sim 2 \times 10^{-2}$), the corrections to the $gg \to Zh$ cross-section are below $5\%$ in the first three bins and become $\sim 30\%$ in the fourth bin and $\sim 60\%$ in the last bin. Given the small number of events in this channel, the new-physics impact on the fit is clearly negligible.

The situation could change in the case in which the flavor-universality assumption is relaxed and the top-coupling modifications are allowed to be larger. In this case a dedicated analysis strategy exploiting other processes sensitive to the top couplings, in addition to the $gg \to Zh$ channel, would be required. We notice, however, that relaxing flavor universality could easily induce large flavor-violating effects, unless flavor symmetries or alignment assumptions are imposed.

\section{Results}
\label{sec.results}

In this section, we present our projection of the bounds on the Wilson coefficients.
We first focus, in Section~\ref{sec:exclusive_fit}, on the $\Ofqtrip$ operator. This choice is justified by the fact that the $Zh$ production channel depends more strongly on the $\Ofqtrip$ operator than on the others, as explained in Section~\ref{sec:interf_terms}, so that a much more stringent bound is expected. Moreover, in many new-physics scenarios we expect the size of the various operators to be comparable (see Section~\ref{sec:power_counting}), in which case the constraints on $\Ofqtrip$ will be the dominant ones.
Afterwards we perform a full fit including the four effective operators, $\Ofqtrip$, $\Ofqsing$, $\Ofu$, and $\Ofd$, quantitatively verifying the statement above.

\subsection{Single operator analysis: $\Ofqtrip$}\label{sec:exclusive_fit}

The 95\% C.L. bounds on $c_{\varphi q}^{(3)}$ from a one-operator fit are given by
\begin{equation}\label{eq:oq3_bounds}
\begin{split}
&[-2.1, 2.0] \times 10^{-3}\quad 1\%\;{\rm syst.}\\
&[-2.6, 2.4] \times 10^{-3}\quad  5\%\;{\rm syst.}\\
& [-3.2, 2.8] \times 10^{-3}\quad  10\%\;{\rm syst.}
\end{split}
\end{equation}
To give an idea of the impact of systematic uncertainties, we considered three benchmark scenarios, with $1\%$, $5\%$, and $10\%$ uncorrelated systematic error in each bin.
We notice that the bounds for positive and negative values of the Wilson coefficient are quite similar. As can be checked from the results
in Tables~\ref{tab:sigma_full} and~\ref{tab:sigma_full_lep}, this is due to the fact that the new-physics contributions are dominated by the linear interference terms with the SM.

Interestingly, the bounds in Eq.~(\ref{eq:oq3_bounds}) are competitive with (and actually slightly better than) the ones obtained with a similar one-operator fit in the $Wh \rightarrow \ell\nu \gamma\gamma$ production channel~\cite{Bishara:2020vix}, which for $5\%$ systematic error gives $c_{\varphi q}^{(3)} \in [-3.3, 2.9] \times 10^{-3}$.
A combination of the two analyses thus allows for a significant improvement in the $\cfqtrip$ bounds, which at 95\% C.L. become
\begin{equation}\label{eq:oq3_bounds_combined}
\begin{split}
&[-1.6, 1.6] \times 10^{-3}\quad 1\%\;{\rm syst.}\\
&[-2.0, 1.9] \times 10^{-3}\quad  5\%\;{\rm syst.}\\
&[-2.4, 2.2] \times 10^{-3}\quad  10\%\;{\rm syst.}
\end{split}
\end{equation}

For comparison, the projections obtained for the leptonic $WZ$ channel at FCC-hh, assuming $5\%$ systematics, give a bound~\cite{Franceschini:2017xkh} 
\begin{equation}
\label{eq:FCChh_bounds}
\begin{array}{l@{\hspace{.5em}}l@{\hspace{2.em}}l@{\hspace{2.em}}l}
    \textrm{FCC-hh} & {(20\;{\rm ab}^{-1})} & c_{\varphi q}^{(3)} \in [-1.8, 1.4] \times 10^{-3}\; & 5\%\;{\rm syst.},
\end{array}
\end{equation}
which is only slightly more stringent than the bound we find combining the $Wh$ and $Zh$ channels. Further combination with the $WZ$ bounds can therefore lead to a significant improvement in the sensitivity.

Other current and projected bounds at 95\% C.L.~on $\chqt$ are:
\begin{equation}\label{eq:bounds_1}
\def\arraystretch{1.3}
\begin{array}{l@{\hspace{.8em}}l@{\hspace{2em}}l}
\textrm{LEP \cite{LEP:2003aa}} &  & [-5.7, 5.7] \times 10^{-1}\;, \\
\textrm{HL-LHC \cite{deBlas:2019rxi,deBlas:2019wgy}} & {(3\;{\rm ab}^{-1})} &  [-3.9, 3.9] \times 10^{-2}\,\left([-0.01, 0.01]\right)\;,  \\
\textrm{HE-LHC \cite{Franceschini:2017xkh}} & {(27\;{\rm TeV},\, 10\;{\rm ab}^{-1})} &  [-4.0, 3.3] \times 10^{-3}\quad
\text{w/5\%\;syst.},\\
\textrm{FCC-ee \cite{deBlas:2019wgy}} &  &  [-6.3, 6.3] \times 10^{-3}\;([-4.8, 4.8] \times 10^{-4})\;, \\
\textrm{CLIC/ILC \cite{deBlas:2019wgy}} &  & [-7.8, 7.8] \times 10^{-3}\;([-6.3, 6.3] \times 10^{-3})\;, \\
\textrm{CEPC \cite{deBlas:2019wgy}} &  & [-9.2, 9.2] \times 10^{-3}\;([-1.1, 1.1] \times 10^{-3})\;.
\end{array}
\end{equation}
We also quote in parentheses the bound from one-operator fits. All these bounds are much weaker than the ones from the $Wh$, $Zh$ and $WZ$ channels, with the exception of the FCC-ee and CEPC one-operator fits.

\begin{table}
\begin{centering}
\begin{tabular}{c|c|c}
\toprule
Coefficient & Profiled Fit & One-Operator Fit \tabularnewline
\midrule
$c_{\varphi q}^{(3)}$ &
\begin{tabular}{ll}
\rule{0pt}{1.25em}$[-5.2,\,3.1]\times10^{-3}$ & $1\%$ syst.\\
\rule{0pt}{1.25em}$[-6.7,\,3.3]\times10^{-3}$ & $5\%$ syst.\\
\rule[-.65em]{0pt}{1.9em}$[-8.2,\,3.7]\times10^{-3}$ & $10\%$ syst.
\end{tabular}
&
\begin{tabular}{ll}
\rule{0pt}{1.25em}$[-2.1,\,2.0]\times10^{-3}$ & $1\%$ syst.\\
\rule{0pt}{1.25em}$[-2.6,\,2.4]\times10^{-3}$ & $5\%$ syst.\\
\rule[-.65em]{0pt}{1.9em}$[-3.2,\,2.8]\times10^{-3}$ & $10\%$ syst.
\end{tabular}
\tabularnewline

\hline 
\begin{tabular}{c}
      $c_{\varphi q}^{(3)}$ \\
      $(+Wh \text{~\cite{Bishara:2020vix}})$
\end{tabular}
&
\begin{tabular}{ll}
\rule{0pt}{1.25em}$[-2.5,\,2.1]\times10^{-3}$ & $1\%$ syst.\\
\rule{0pt}{1.25em}$[-3.0,\,2.4]\times10^{-3}$ & $5\%$ syst.\\
\rule[-.65em]{0pt}{1.9em}$[-3.7,\,2.7]\times10^{-3}$ & $10\%$ syst.
\end{tabular}
&
\begin{tabular}{ll}
\rule{0pt}{1.25em}$[-1.6,\,1.6]\times10^{-3}$ & $1\%$ syst.\\
\rule{0pt}{1.25em}$[-2.0,\,1.9]\times10^{-3}$ & $5\%$ syst.\\
\rule[-.65em]{0pt}{1.9em}$[-2.4,\,2.2]\times10^{-3}$ & $10\%$ syst.
\end{tabular}
\tabularnewline

\hline
$c_{\varphi q}^{(1)}$ &
\begin{tabular}{ll}
\rule{0pt}{1.25em}$[-1.3,\,1.4]\times10^{-2}$ & $1\%$ syst.\\
\rule{0pt}{1.25em}$[-1.5,\,1.5]\times10^{-2}$ & $5\%$ syst.\\
\rule[-.65em]{0pt}{1.9em}$[-1.6,\,1.5]\times10^{-2}$ & $10\%$ syst.
\end{tabular}
&
\begin{tabular}{ll}
\rule{0pt}{1.25em}$[-1.1,\,1.2]\times10^{-2}$ & $1\%$ syst.\\
\rule{0pt}{1.25em}$[-1.2,\,1.2]\times10^{-2}$ & $5\%$ syst.\\
\rule[-.65em]{0pt}{1.9em}$[-1.2,\,1.2]\times10^{-2}$ & $10\%$ syst.
\end{tabular}
\tabularnewline

\hline
$c_{\varphi u}$ &
\begin{tabular}{ll}
\rule{0pt}{1.25em}$[-2.0,\,1.6]\times10^{-2}$ & $1\%$ syst.\\
\rule{0pt}{1.25em}$[-2.1,\,1.7]\times10^{-2}$ & $5\%$ syst.\\
\rule[-.65em]{0pt}{1.9em}$[-2.3,\,1.8]\times10^{-2}$ & $10\%$ syst.
\end{tabular}
&
\begin{tabular}{ll}
\rule{0pt}{1.25em}$[-1.9,\,0.89]\times10^{-2}$ & $1\%$ syst.\\
\rule{0pt}{1.25em}$[-2.1,\,0.96]\times10^{-2}$ & $5\%$ syst.\\
\rule[-.65em]{0pt}{1.9em}$[-2.2,\,1.0]\times10^{-2}$ & $10\%$ syst.\\
\end{tabular}
\tabularnewline

\hline
$c_{\varphi d}$ &
\begin{tabular}{ll}
\rule{0pt}{1.25em}$[-2.1,\,2.3]\times10^{-2}$ & $1\%$ syst.\\
\rule{0pt}{1.25em}$[-2.2,\,2.4]\times10^{-2}$ & $5\%$ syst.\\
\rule[-.65em]{0pt}{1.9em}$[-2.3,\,2.5]\times10^{-2}$ & $10\%$ syst.
\end{tabular}
&
\begin{tabular}{ll}
\rule{0pt}{1.25em}$[-1.4,\,2.2]\times10^{-2}$ & $1\%$ syst.\\
\rule{0pt}{1.25em}$[-1.5,\,2.2]\times10^{-2}$ & $5\%$ syst.\\
\rule[-.65em]{0pt}{1.9em}$[-1.5,\,2.2]\times10^{-2}$ & $10\%$ syst.\\
\end{tabular} \\
\bottomrule

\end{tabular}
\par\end{centering}
\caption[caption]{Bounds at $95\%$ C.L.~on the coefficients of the $\Ohqt$, $\Ohq$, $\Ohu$ and $\Ohd$ operators (with the normalization $\Lambda = 1 \, \text{TeV}$). The row labelled as $c_{\varphi q}^{(3)}\,(+Wh \text{~\cite{Bishara:2020vix}})$ provides the bounds obtained from the combination of our analysis with the one for the $Wh$ channel presented in Ref.~\cite{Bishara:2020vix}. 
{\bf Left column:} Bounds from the global fit, profiled over the other coefficients. {\bf Right column:} Bounds from a one-operator fit (i.e. setting the other coefficients to zero).
}
\label{tab:bounds_summary}
\end{table}

\subsection{Full analysis}\label{sec:global_fit}

The bounds from the four-operator fit, i.e, with $\Ofqtrip$, $\Ofqsing$, $\Ofu$ and $\Ofd$, profiled over three operators at a time are reported in the column labeled ``Profiled Fit'' in Table~\ref{tab:bounds_summary}.
These results clearly show that, even in a full fit, the sensitivity to the $\Ofqtrip$ operator is much higher than to the other operators.

The row labelled $\cfqtrip (+Wh)$ in Table~\ref{tab:bounds_summary}
corresponds to the combination with the bounds from the one-operator fit in the $Wh$ channel presented in Ref.~\cite{Bishara:2020vix}. Notice that, in the $Wh$ channel, the operator $\Ofqtrip$ is the only one that grows quadratically with the energy, which justifies our choice of the one-operator fit. This combination leads to an improvement by a factor $\sim 2$ in the bounds on $\cfqtrip$, whereas the impact on the determination of the other Wilson coefficients is negligible (and not reported in the table).
This is because the bounds on the various operators in the fit are nearly uncorrelated.
This feature is confirmed by the fact that the bounds from single-operator fits on $\cfqsing$, $\cfu$ and $\cfd$ are nearly equal to the ones coming from the global fit
(see ``One-Operator Fit'' column in Table~\ref{tab:bounds_summary}). On the contrary, the
determination of $\cfqtrip$ gets significantly weaker in the global fit than in the single-operator analysis, since large values of the other Wilson coefficients can easily compensate the linear corrections in $\cfqtrip$.

\begin{figure}[t]
	\centering
	\includegraphics[width=0.47\linewidth]{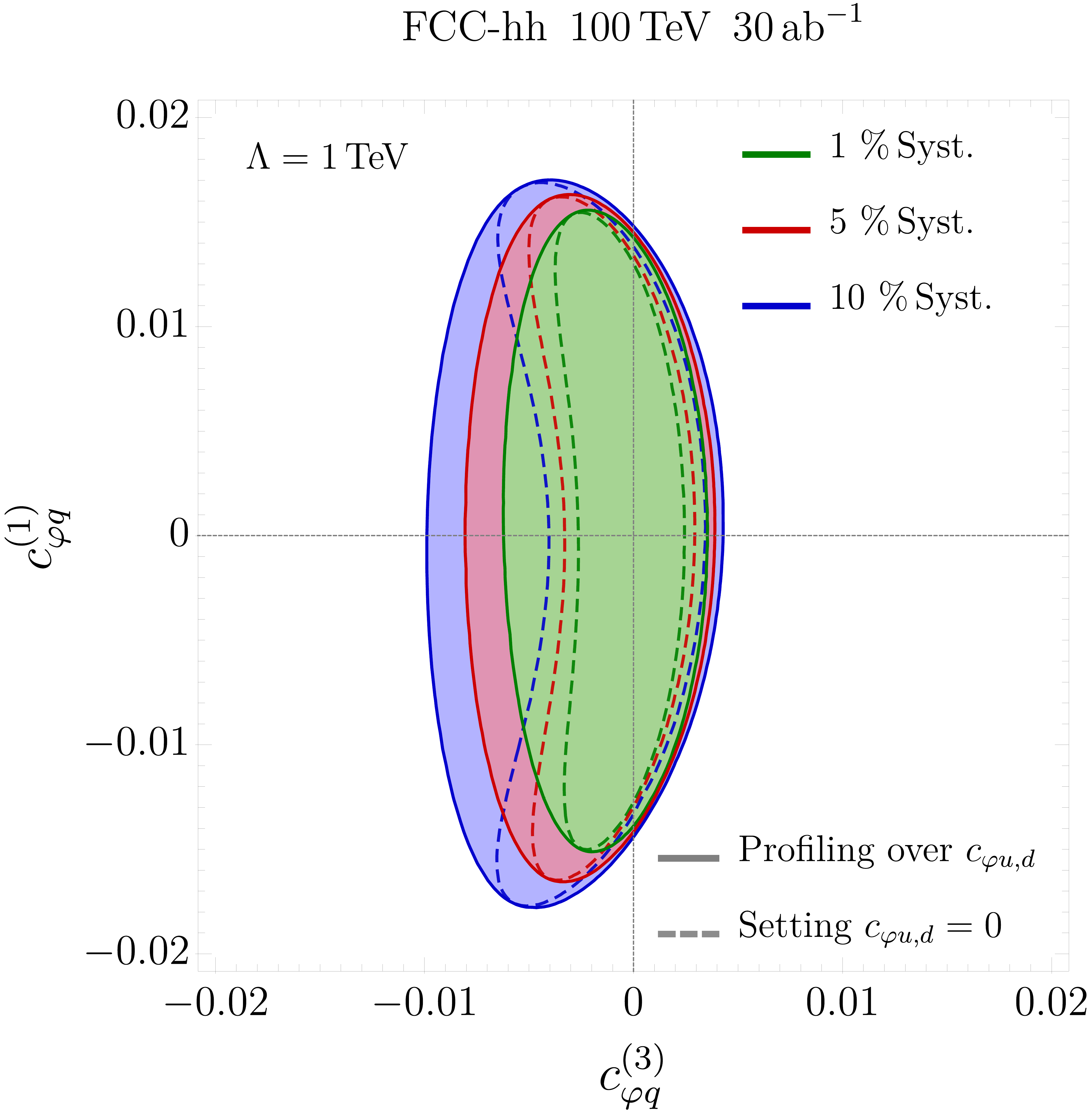}
	\hfill
	\includegraphics[width=0.47\linewidth]{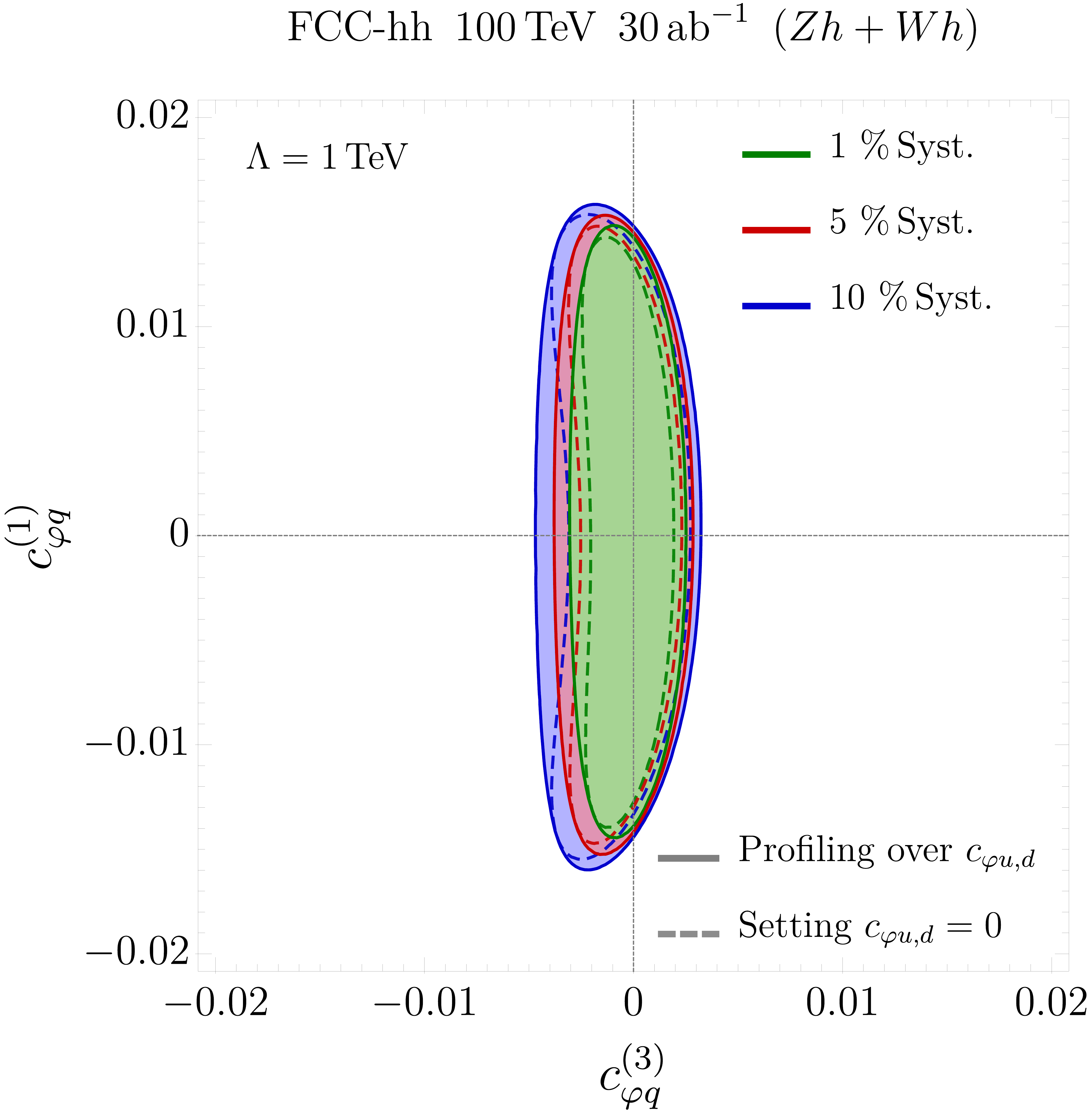}
	\caption{Expected $95\%$ C.L. bounds on $\chq$ and $\chqt$ at the FCC-hh.
	Bounds in green, red, blue, assume $1\%$, $5\%$ and $10\%$ systematic error.
	Solid (dashed) lines correspond to the bounds when profiling over (setting to zero) the Wilson coefficients not appearing in the plot.
	\textbf{Left panel:} Bounds obtained from the $Zh$ channels analysis presented in this paper.
	\textbf{Right panel:} Bounds obtained from the combination with the $Wh$ analysis in Ref.~\cite{Bishara:2020vix}.
	}
	\label{fig:2D_bounds}
\end{figure}

In Fig.~\ref{fig:2D_bounds}, we provide the fits in the $(\cfqtrip, \cfqsing)$ plane, obtained either by profiling (solid lines) or by setting to zero the $\cfu$
and $\cfd$ operators (dashed lines). In the latter case some correlation in the fit (induced by the mixed $\cfqtrip - \cfqsing$ terms) can be seen, which however disappears in the profiled fit. In the right panel of the figure we also show how the fit is modified by combining with the $Wh$ analysis of Ref.~\cite{Bishara:2020vix}. The main impact of the combination is an improvement of the bounds on $\cfqtrip$, whereas
the bound along the $\cfqsing$ direction is nearly unaffected.

Finally we can compare the bounds we obtained with the expected sensitivity at the LHC and other proposed future colliders (our fits always implicitly also rely on LEP measurements that are crucial in particular to size the SM input parameters). The current and projected 95\% C.L. bounds on $\chq$ from a global analysis are ~\cite{Ellis:2018gqa,deBlas:2019wgy,deBlas:2019rxi}:
\begin{equation}
\begin{array}{l@{\hspace{.8em}}l@{\hspace{3.em}}l@{\hspace{2.em}}l}
\textrm{LHC } \ &{\text{Run 2 data}}& [-0.132, 0.066]\,,  \\
\textrm{HL-LHC } \ &{(3\;{\rm ab}^{-1})} &[-0.085, 0.085]\,\left([-0.03, 0.03]\right)\,,  \\
\textrm{CLIC/ILC }& & [-0.07, 0.07]\,\left([-0.03, 0.03]\right)\,,\\
\textrm{CEPC }& & [-0.008, 0.008]\,\left([-0.003, 0.003]\right)\,,\\
\textrm{FCC-ee }& & [-0.018, 0.018]\,\left([-0.0017, 0.0017]\right)\,,\\
\end{array}
\end{equation}
for $\Lambda = 1$ TeV. The current and projected bounds on $\chu$ are ~\cite{Ellis:2018gqa,deBlas:2019wgy,deBlas:2019rxi}:
\begin{equation}
\begin{array}{l@{\hspace{.8em}}l@{\hspace{3.em}}l@{\hspace{2.em}}l}
\textrm{LHC } \ &{\text{Run 2 data}}&[-0.36, 0.36]\,,  \\
\textrm{HL-LHC } \ &{(3\;{\rm ab}^{-1})}&[-0.24, 0.24]\,\left([-0.06, 0.06]\right)\,,  \\
\textrm{CLIC/ILC }& &[-0.17, 0.17]\,\left([-0.07, 0.07]\right)\,,\\
\textrm{CEPC }& &[-0.02, 0.02]\,\left([-0.007, 0.007]\right)\,,\\
\textrm{FCC-ee }& & [-0.04, 0.04]\,\left([-0.003, 0.003]\right)\,.\quad\quad\quad\\
\end{array}
\end{equation}
Finally the bounds on $\chd$ are ~\cite{Ellis:2018gqa,deBlas:2019wgy,deBlas:2019rxi}:
\begin{equation}
\begin{array}{l@{\hspace{.8em}}l@{\hspace{3.em}}l@{\hspace{2.em}}l}
\textrm{LHC } \ &{\text{Run 2 data}}& [-0.62, 0.50]\,,  \\
\textrm{HL-LHC } \ &{(3\;{\rm ab}^{-1})}&[-0.45, 0.45]\,\left([-0.09, 0.09]\right)\,,  \\
\textrm{CLIC/ILC }& & [-0.4, 0.4]\,\left([-0.1, 0.1]\right)\,,\\
\textrm{CEPC }& & [-0.04, 0.04]\,\left([-0.009, 0.009]\right)\,,\\
\textrm{FCC-ee }& &[-0.095, 0.095]\,\left([-0.004, 0.004]\right)\,. \quad\quad\\
\end{array}
\end{equation}

\begin{figure}[t]
	\centering
	\includegraphics[width=0.485\linewidth]{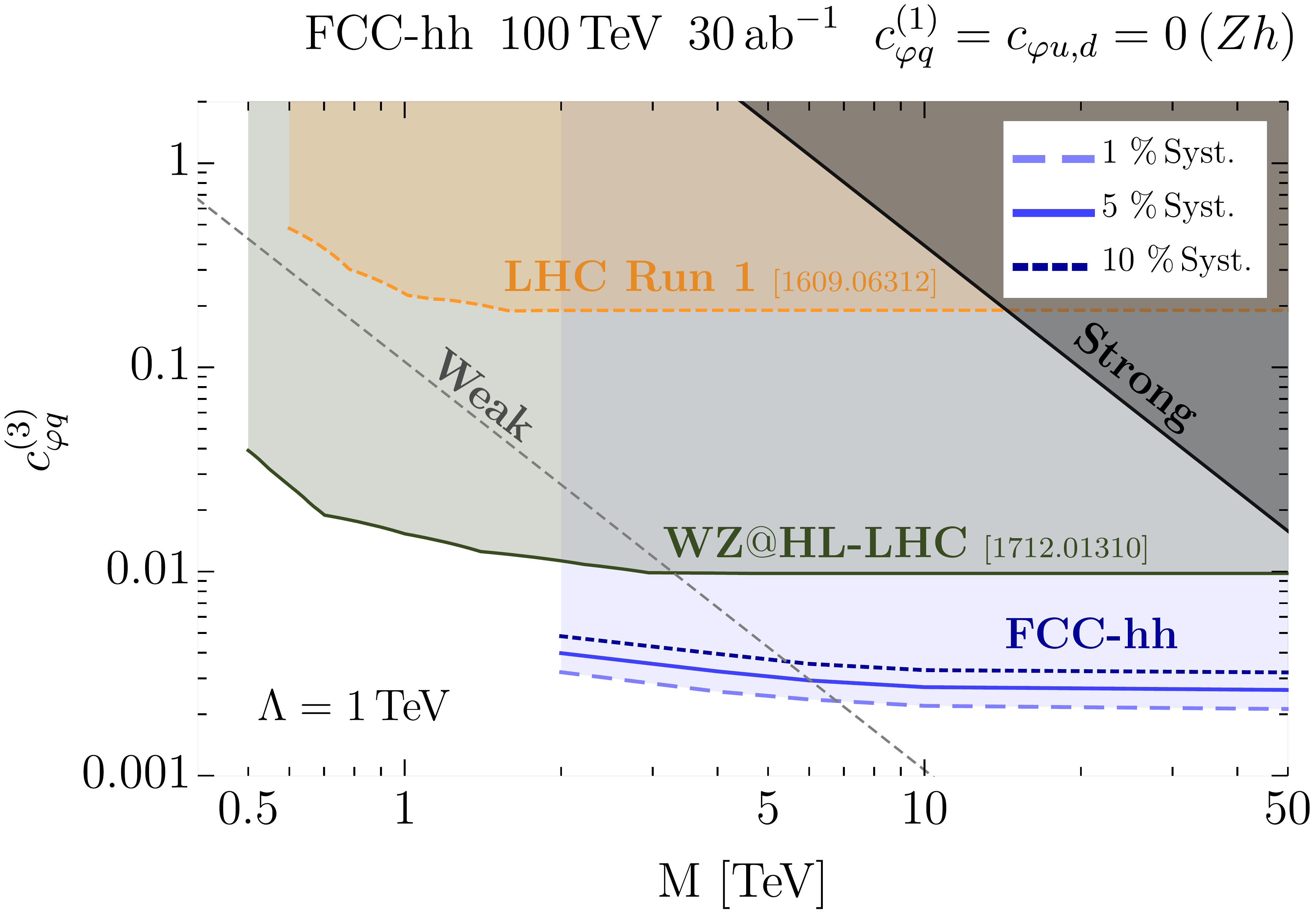} \hfill
	\includegraphics[width=0.485\linewidth]{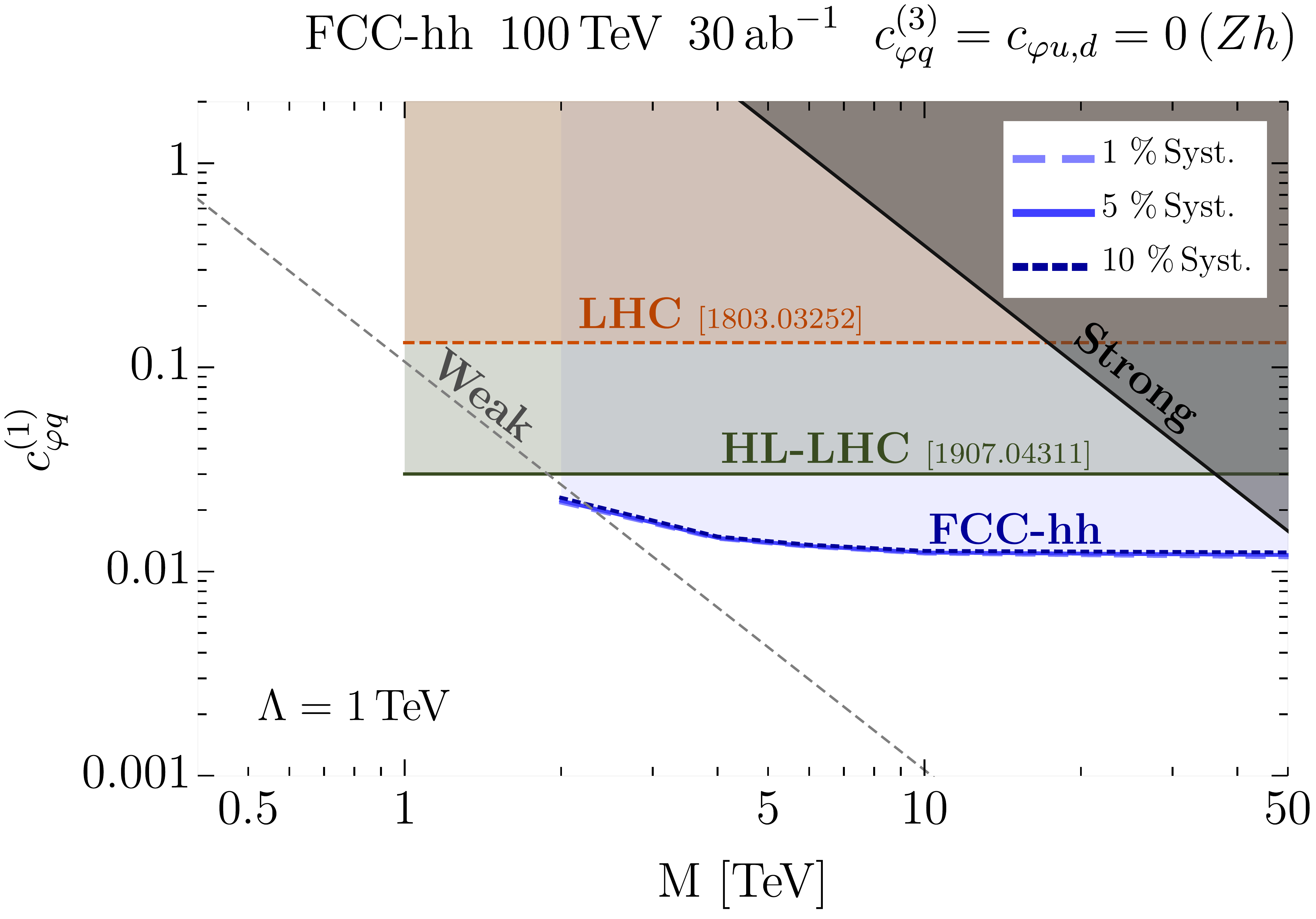} \\
	\vspace{0.3cm}
    \includegraphics[width=0.485\linewidth]{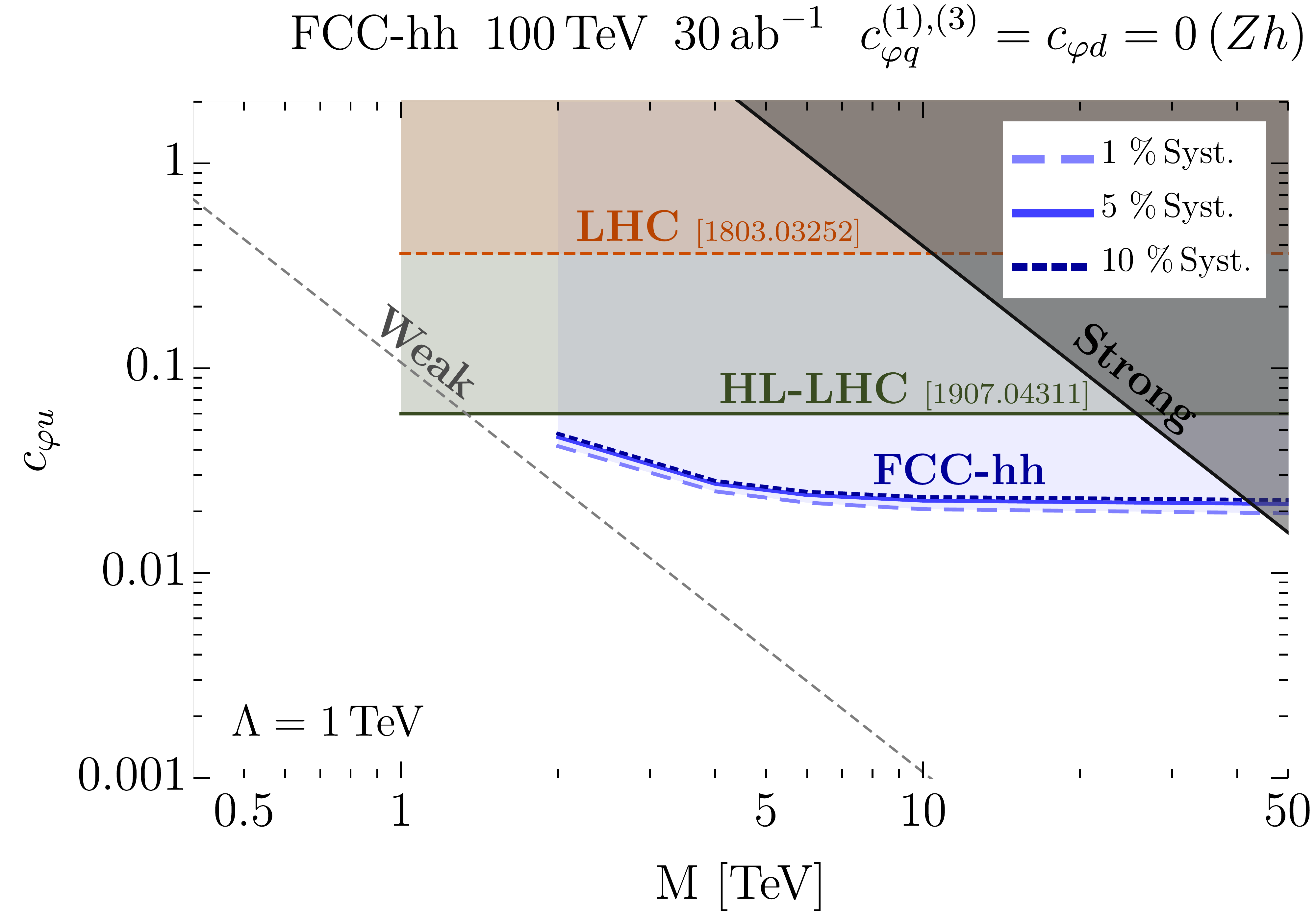}\hfill
	\includegraphics[width=0.485\linewidth]{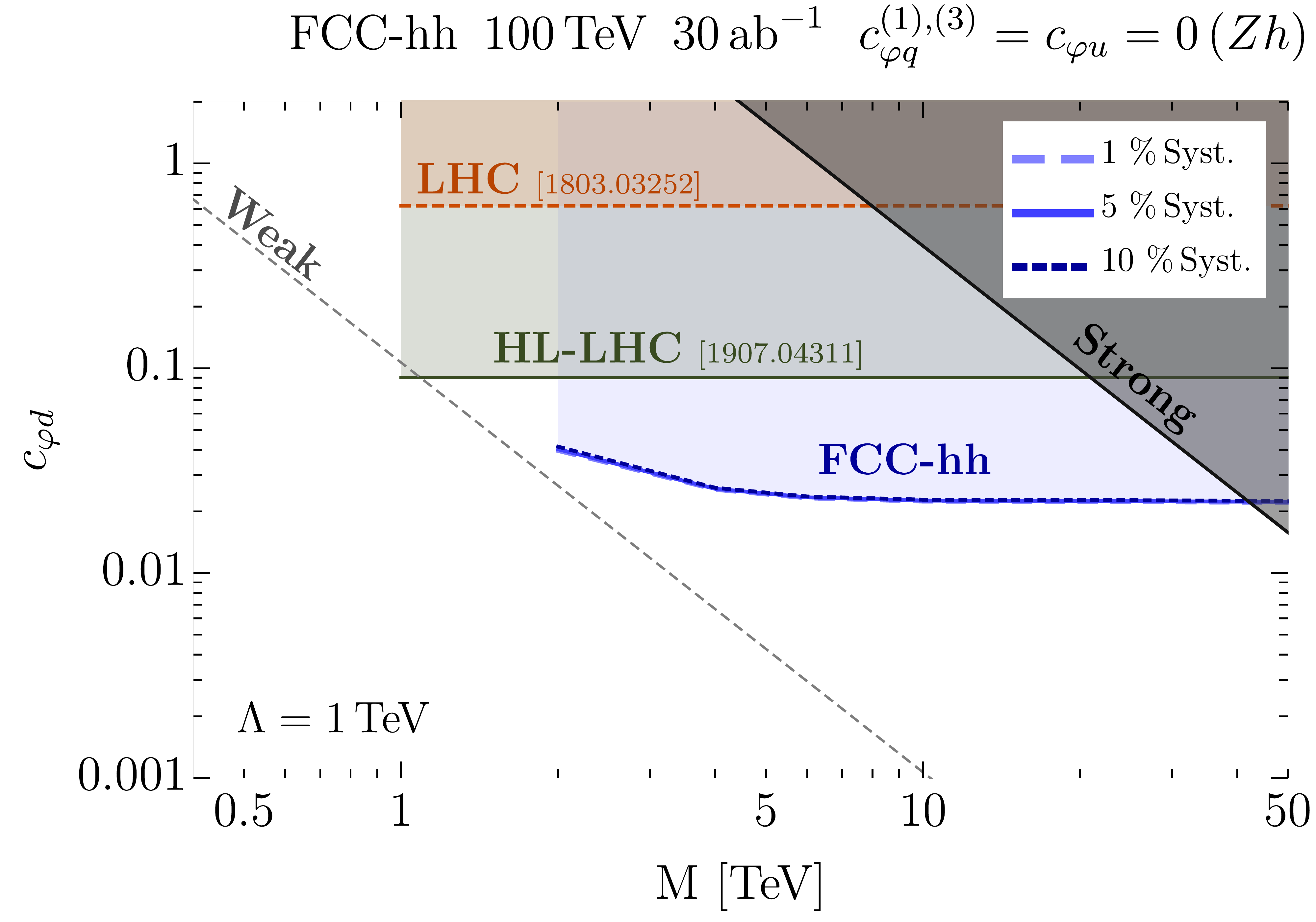}
	\caption{ Bounds on $c_{\varphi q}^{(3)}$, $\chq$, $\chu$ and $\chd$ from one-operator fits
	as functions of the maximal-invariant-mass cut $M$. The bounds correspond to $95\%$ C.L. ($\Delta \chi^2 = 3.84$). 
	The dashed, solid and dotted lines show the bounds for $1\%$, $5\%$ and $10\%$ systematic errors. The current bounds and the projections for some future hadron colliders are also shown.
	For $\chqt$, we show the LHC Run 1 bound from Ref. \cite{Falkowski:2016cxu} and the projections from the $WZ$ channel at HL-LHC from Ref. \cite{Franceschini:2017xkh}. For the rest of operators, we show the global LHC data fit bound from Ref. \cite{Ellis:2018gqa} and the 1-operator fit at HL-LHC from Ref. \cite{deBlas:2019wgy}. LHC bounds are shown in orange and HL-LHC ones are shown in dark green. The dashed gray and solid black lines show the values of the Wilson coefficient expected in weakly-coupled ($c \sim g^2/(4 M^2)$) and strongly-coupled ($c \sim (2\pi)^2/(M^2)$) new physics models~\cite{Franceschini:2017xkh}.
	}
	\label{fig:fit_cphiq}
\end{figure}

We see that, for all operators, our analysis provides bounds that are competitive with the ones expected from global fits at other future colliders. On the other hand, if one-operator fits are considered, FCC-ee and CEPC will surpass our bounds on all four operators by roughly one order of magnitude. 

The bounds from single-operator fits as functions of the maximal invariant mass of the events used in the analysis are shown in Fig.~\ref{fig:fit_cphiq}. This kind of analysis is very useful in testing the validity of the EFT description and understanding what kind of theories can be tested. We see that, for all the operators, the bounds saturate for $M \sim 5\;\textrm{TeV}$, signalling that our bounds are valid for a cut-off $\Lambda \gtrsim 5\; \textrm{TeV}$. One can also appreciate the fact that only the bounds on the $\Ofqtrip$ operator are strong enough to test the region of weakly-coupled new physics, whereas the other operators can mostly test theories in which the new dynamics is strongly coupled.

\subsection{Connection to aTGCs and Universal Theories}
\label{sec:Higgs_EDM_aTGC}
So far we used our analysis to set bounds on the $\Ohqt, \, \Ohq, \, \Ohu$ and $\Ohd$ operators in the Warsaw basis. Nonetheless, for certain classes of UV models it can be convenient to interpret the bounds in terms of anomalous Triple Gauge Couplings (aTGC).
The $\Ohqt, \, \Ohq, \, \Ohu$ and $\Ohd$ operators can be rewritten in terms of vertex corrections, $\delta g_{L,R}^{Zq}$, and aTGC, $\delta g_{1z}$ and $\delta\kappa_{\gamma}$ by adopting the Higgs basis~\cite{Gupta:2014rxa,deFlorian:2016spz}:
\begin{eqnarray}
\begin{aligned}
    \chqt = & +   \frac{\Lambda^2}{4 m_W^2} g^2 \left(\delta g_L^{Zu} - \delta g_L^{Zd} - c_{\textsc w}^2 \, \delta g_{1z}\right) \\
    \chq = & - \frac{\Lambda^2}{4 m_W^2} g^2\left(\delta g_L^{Zu} + \delta g_L^{Zd} +\frac{1}{3}\left(t_{\textsc w}^2 \delta\kappa_{\gamma}-s_{\textsc w}^2 \delta g_{1z}\right)\right) \\
    \chu = & - \frac{\Lambda^2}{2 m_W^2} g^2 \left(\delta g_{R}^{Zu}+\frac{2}{3}\left(t_{\textsc w}^2 \delta\kappa_{\gamma}-s_{\textsc w}^2 \delta g_{1z}\right)\right) \\ 
    \chd = & - \frac{\Lambda^2}{2 m_W^2} g^2 \left(\delta g_{R}^{Zd}-\frac{1}{3}\left(t_{\textsc w}^2 \delta\kappa_{\gamma}-s_{\textsc w}^2 \delta g_{1z}\right)\right) \\
    \end{aligned}
    \label{eq:Higgsbasis}
\end{eqnarray}
where $c_{\textsc w}$, $s_{\textsc w}$ and $t_{\textsc w}$  are the cosine, sine and tangent of the weak mixing angle respectively. From Eq.~\eqref{eq:Higgsbasis} we see that if the $\delta g_{L,R}^{Zq}$ coefficients are negligible, one can recast our diboson bounds on $\chqt,\, \chq, \, \chu, \, \chd$ to constraints on $\delta g_{1z}$ and $\delta\kappa_{\gamma}$. This is particularly justified for universal theories where $\delta g_{L,R}^{Zq}$ correspond to combinations of the oblique parameters S, T, W and Y ~\cite{Franceschini:2017xkh, Grojean:2018dqj}, which are expected to be highly constrained through various measurements at FCC-ee and FCC-hh.
With these assumptions, the combined analysis of the $Zh$ and $Wh$ channels gives the following constraints (for $5\%$ systematic uncertainty)
\begin{eqnarray}\label{eq:TGCbound_exclusive}
\delta g_{1z} &\in& [-1.8, 2.4] \times 10^{-4} \hspace{1.5em} ([-1.5, 1.7]\times 10^{-4})\,, \\
\delta\kappa_{\gamma} &\in& [-1.5, 2.8] \times 10^{-3} \hspace{1.5em} ([-1.2, 2.8] \times 10^{-3})\,,
\end{eqnarray}
where the bounds in parenthesis are obtained through one-operator fits.

\begin{figure}[t]
	\centering
	\includegraphics[width=0.6\linewidth]{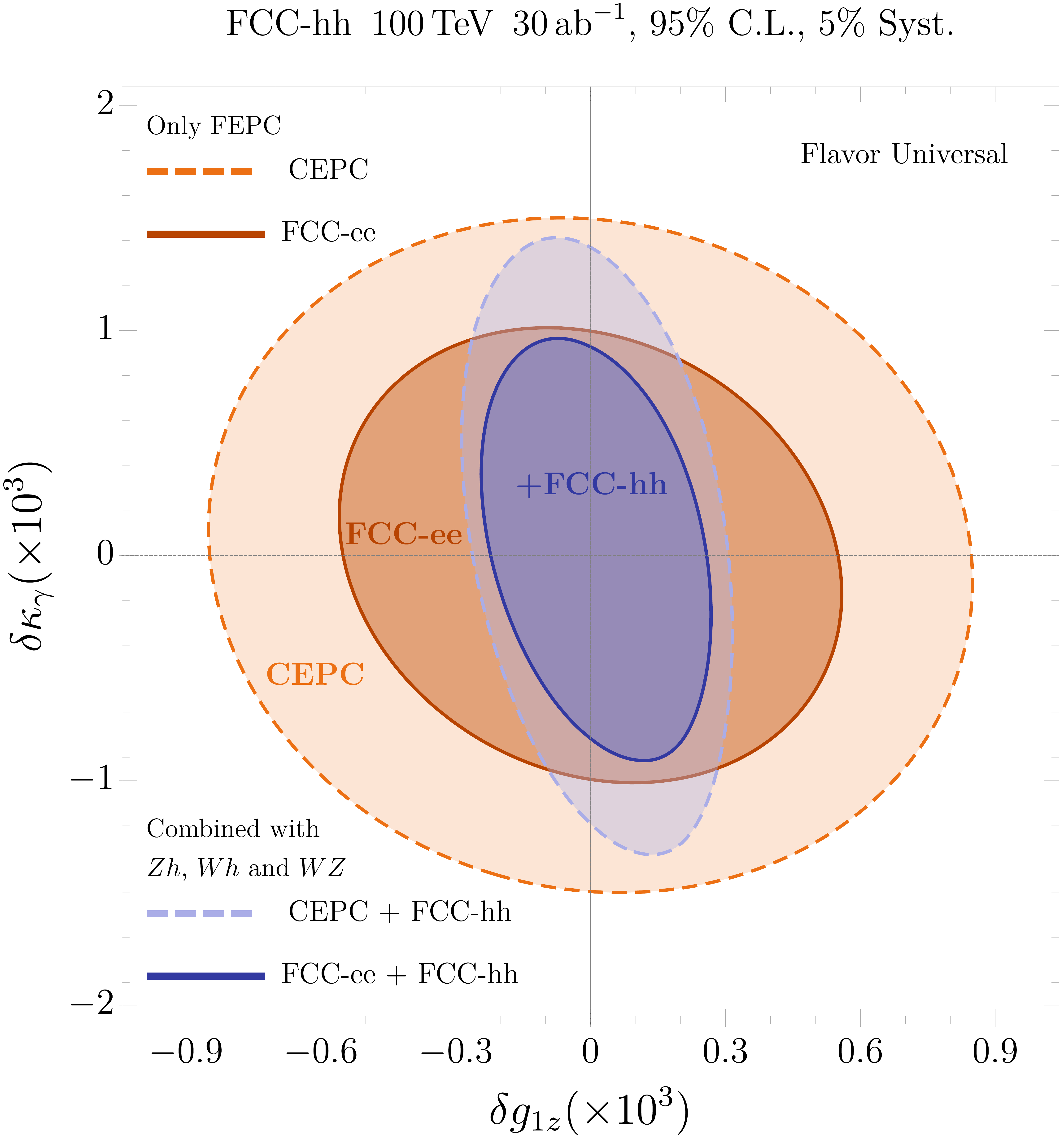}
	\caption{Projected 95\% C.L. bounds on the anomalous Triple Gauge Couplings $\delta \kappa_\gamma \,, \delta g_{1z}$ for flavor universal theories at future colliders.
	In dark and light orange, we show the expected bounds at FCC-ee and CEPC~\cite{deBlas:2019wgy}. In dark (light) blue, we show the combination of the FCC-ee (CEPC) fit with the projections at FCC-hh for the diboson channels $WZ \to \ell \nu \ell^{+}\ell^{-}$~\cite{Franceschini:2017xkh}, $Wh \to \ell \nu \gamma \gamma$~\cite{Bishara:2020vix}, and the combined $Zh \to \ell^+ \ell^- \gamma \gamma$ and $Zh \to \nu \bar{\nu}  \gamma \gamma$ (this work).
	In the FCC-hh projections we assume $5\%$ systematic uncertainty.
	}
	\label{fig:universal}
\end{figure}

Another way to estimate the impact of diboson searches at FCC-hh is to compare their reach with the sensitivity at future lepton colliders. As an illustrative example we show, in Fig.~\ref{fig:universal}, the expected bounds on the aTGC parameters $\delta g_{1z}$ and $\delta \kappa_\gamma$ for CEPC and FCC-ee. These bounds are obtained through a global fit, which takes into account 18 effective operators in flavor-universal
theories~\cite{deBlas:2019wgy}.\footnote{When flavor-universal theories are mapped into the 18 operators used in the fit, an ambiguity in the choice of the 4-lepton effective interactions is present. To obtain the results shown in Fig.~\ref{fig:universal} we used the choice $[c_{\ell\ell}]_{1221} = 0$. We however checked that the choice of $[c_{\ell\ell}]_{1221}$ only has a mild impact on the bounds.}
We also show in the same plot the impact of the combination of the CEPC and FCC-ee constraints with the FCC-hh fit of diboson channels. For the FCC-hh fit we combine the projections for $WZ \to \ell \nu \ell^{+}\ell^{-}$~\cite{Franceschini:2017xkh}, $Wh \to \ell \nu \gamma \gamma$~\cite{Bishara:2020vix}, and the two channels $Zh \to \ell^+ \ell^- \gamma \gamma$ and $Zh \to \nu \bar{\nu} \gamma \gamma$ (this work), and we assume $5\%$ systematic uncertainty.

We find that, once the diboson channels at FCC-hh are included, the projected bounds on $\delta g_{1z}$ improve by a factor $\sim 2$ (3) with respect to the bounds at FCC-ee (CEPC). On the other hand the bounds on $\delta \kappa_\gamma$ are only marginally affected. This behavior is due to the fact that diboson channels have a very good sensitivity to $\cfqtrip$, which only depends on $\delta g_{1z}$ (see Eq.~(\ref{eq:Higgsbasis})), while information on $\delta \kappa_\gamma$ can only be extracted from $\cfqsing$, $\cfu$ and $\cfd$, whose expected bounds are much weaker.

For completeness we report in the following the allowed $95\%$ C.L.~regions for $\delta g_{1z}$,
\begin{equation}
\begin{array}{l@{\hspace{.8em}}l@{\hspace{3.em}}l@{\hspace{2.em}}l}
\textrm{FCC-ee } \ & & [-4.5, 4.5] \times 10^{-4}\,,  \\
\textrm{FCC-ee $+$ Diboson at FCC-hh} \ & &[-2.0, 2.2] \times 10^{-4}\,,  \\
\textrm{CEPC }& & [-6.8, 6.8]\times 10^{-4}\,,\\
\textrm{CEPC $+$ Diboson at FCC-hh}& & [-2.3, 2.5]\times 10^{-4} \,,\\
\end{array}
\label{eq:BoundsDg1z_FlaUniv}
\end{equation}
and for $\delta \kappa_\gamma$,
\begin{equation}
\begin{array}{l@{\hspace{.8em}}l@{\hspace{3.em}}l@{\hspace{2.em}}l}
\textrm{FCC-ee } \ & & [-8.1, 8.1] \times 10^{-4}\,,  \\
\textrm{FCC-ee $+$ Diboson at FCC-hh} \ & &[-7.4, 7.7] \times 10^{-4}\,,  \\
\textrm{CEPC }& & [-1.2, 1.2]\times 10^{-3}\,,\\
\textrm{CEPC $+$ Diboson at FCC-hh}& & [-1.1, 1.1]\times 10^{-3} \,.\\
\end{array}
\label{eq:BoundsDka_FlaUniv}
\end{equation}
These bounds are obtained from the global fit profiling over all other parameters.

\section{Summary and conclusions} \label{sec:conclusions}

The next generation of hadron colliders, thanks to the enhanced cross-sections and the high integrated luminosity,
will allow us to revisit several electroweak processes in a cleaner environment, bringing advantages for precision measurements.
An interesting example is $Vh$ production, which can only be studied at HL-LHC through the $h\rightarrow b \bar{b}$ decay, but becomes accessible at FCC-hh also in the very clean channel $h\rightarrow \gamma\gamma$.
In a previous paper~\cite{Bishara:2020vix}, we exploited the latter channel in $Wh$ production, showing how it can be used to test deviations in the $W$-boson couplings to quarks with high accuracy. In the present work, we extended the analysis to a closely related channel, $Zh$ production, considering the $h\rightarrow \gamma\gamma$ decay channel together with a $Z$-boson decay into a pair of charged leptons or neutrinos. In spite of its smaller cross-section, the $Zh$ channel is extremely interesting since it can be used to probe a larger set of new-physics effects, including deviations in the $Z$ couplings to quarks that cannot be tested in the $Wh$ channel. 
In the SMEFT framework, and focusing on dimension-6 operators that induce a $\hat{s}$ growth in the amplitude, the $Zh$ production process is sensitive to four ``primary'' operators~\cite{Franceschini:2017xkh}, which in the Warsaw basis correspond to $\Ohqt$, $\Ohq$, $\Ohu$, and $\Ohd$.

Binning in transverse momentum and combining both aforementioned $Z$ decay channels allowed us to obtain a sensitivity on $\Ohqt$ that is competitive with other processes that have higher cross-sections, like $Wh$. Our estimates show that the bounds at FCC-hh can significantly surpass the precision achievable at HL-LHC and at FCC-ee. Furthermore, the combination of the $Zh$ and $Wh$ channels can improve the bound by roughly $20\%$ and further improvement could be achieved via a combination with the $WZ$ channel. This shows the importance of a comprehensive study of all processes available at future colliders in order to correctly assess their potential.

The sensitivity to the $\Ohq$ operator turns out to be significantly smaller, due to a (partially accidental) cancellation between the contributions from up-type and down-type quarks.
We showed that this cancellation
can be partially overcome by implementing a second binning in the rapidity of the $Zh$ system (or the rapidity of the Higgs boson when the former cannot be computed). This double binning exploits the differences in the rapidity distribution due to the parton distribution functions of the various generation-level quarks. Although the improvement is limited in the final state we considered in this paper, this strategy could be useful for final states with higher cross-section (for instance the $h \to b\bar{b}$ channel) or for analogous processes. We also stress that the pattern of deviations in the rapidity distribution depends on the flavor structure of the new-physics effects (see Fig.~\ref{fig:y_binning}), thus it could potentially be a way to disentangle different flavor hypotheses. The sensitivity we get to the $\Ohq$ operator is one order of magnitude better than the one at HL-LHC and is competitive with the one achievable at future lepton colliders (CLIC/ILC, CEPC and FCC-ee).

Finally, our sensitivity to $\Ohu$ and $\Ohd$ is limited by inherent characteristics of the operators, whose contributions only interfere with the SM amplitudes due to the relatively small couplings of the right-handed quarks with the $Z$ boson. The expected bounds are more than one order of magnitude better than the ones at HL-LHC and competitive with the ones from global fits at future lepton colliders.
Regarding the bounds on $\Ohu$ and $\Ohd$, we note that, due to the suppression of the interference with the SM amplitudes, the constraints are mostly driven by the square of the BSM contributions. This might cause some limitation in their interpretability within the EFT formalism.

\begin{figure}[t]
 	\centering
 	\includegraphics[width=0.8\linewidth]{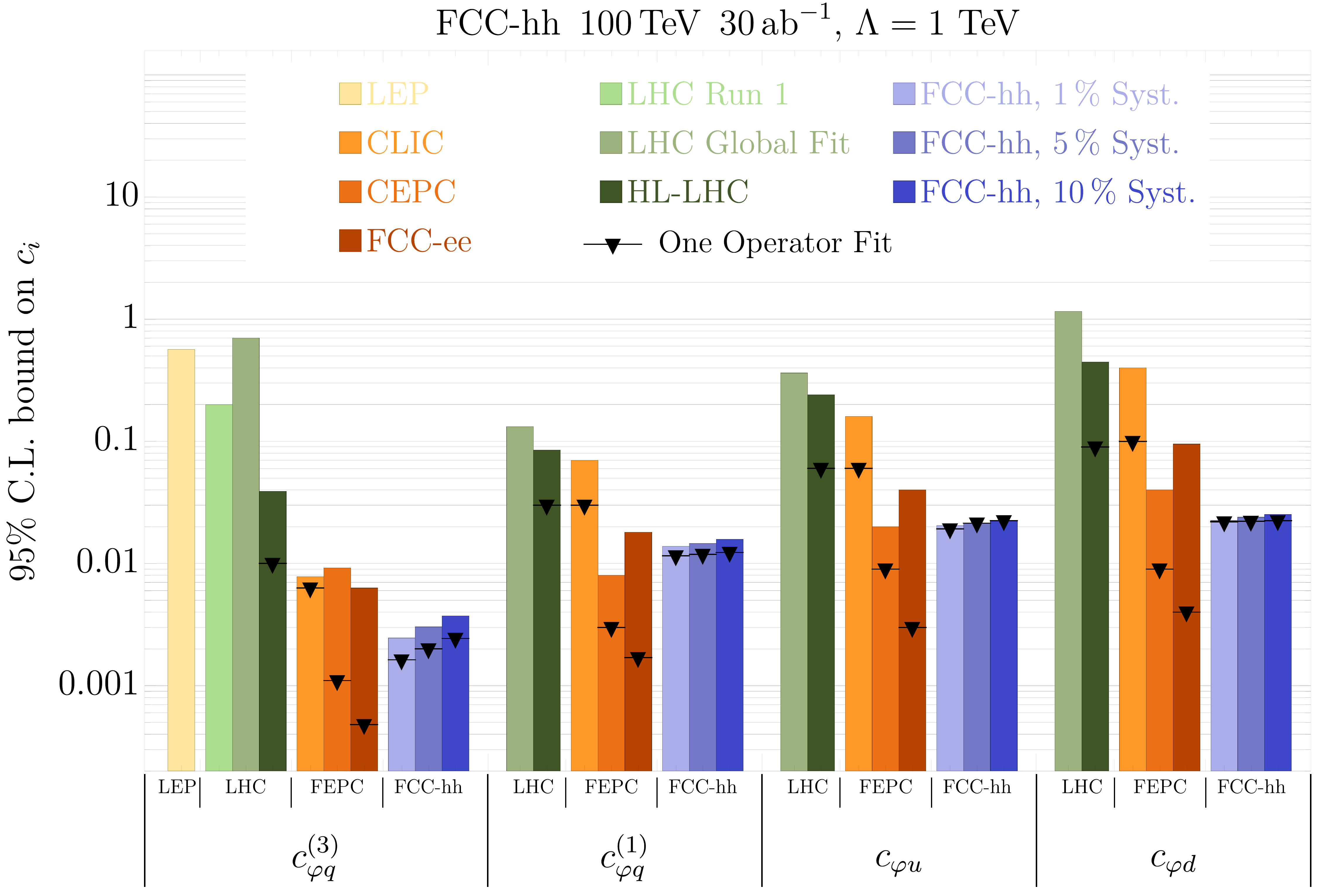}
 	\caption{$95\%$ C.L. bounds on $\chqt$, $\chq$, $\chu$ and $\chd$. In blue, our combined bounds from $Zh\rightarrow \left(\nu\bar{\nu}/\ell^{+}\ell^{-}\right)\gamma\gamma$ and $Wh\rightarrow \ell\nu\gamma\gamma$  at FCC-hh with 30 ab$^{-1}$ for different systematics and computed from a four operator fit. In all cases, the black lines with a triangle on top represent the bound from a one-operator fit instead. In light yellow, the current LEP \cite{LEP:2003aa} bound for $\chqt$. In light green for $\chqt$, the run-1 LHC \cite{Falkowski:2016cxu} bounds. In medium green, the current bound on all the operators from a global fit \cite{Ellis:2018gqa}. In dark green, the projections from a global fit at HL-LHC\cite{deBlas:2019rxi,deBlas:2019wgy}.
 	In light, medium and dark orange, the projected bounds on the operators from a global fit at CLIC, CEPC and FCC-ee respectively \cite{deBlas:2019wgy}.   FEPC stands for Future Electron-Positron Colliders.
 	}
 	\label{fig:comparison_bounds}
 \end{figure}
 
A summary of the projected $95\%$ C.L. bounds on the four operators we considered is shown in Fig.~\ref{fig:comparison_bounds}.
The blue bars correspond to the constraints derived from the profiling of a four-operator fit. On the other hand, the horizontal bars with a triangle indicate the bound obtained from a fit including one operator at a time.
In both fits, we considered three possible values for the systematic uncertainties: $1\%$ (lighter shading), $5\%$ (medium shading), and $10\%$ (darker shading). The systematic uncertainty has a sizeable effect only on the bound for $\Ohqt$.
The $5\%$ scenario is comparable to the present LHC systematics for similar processes, therefore it could be considered as a conservative estimate, while the $10\%$ benchmark is most probably an over-pessimistic one.

Many directions in the assessment of the precision-measurement potential of future hadron colliders could still be explored. Regarding the Higgs-associated production channels ($Wh$ and $Zh$), an interesting direction to follow is the study of the hadronic decay channels, in particular the $h \to b\bar b$ decay which can offer a boost in cross-section, however at the price of significantly larger backgrounds.
Hadronic decays of the $W$ and $Z$ bosons could also be considered.
In the broader context of diboson channels, the $WZ$ and $WW$ production processes can also be exploited to test a similar set of new-physics effects. The former channel has been so far investigated mainly in the fully leptonic final state, while other decays could also be worth considering. $WW$ production is instead much more challenging and very few studies are already available.
Finally, we stress that such analyses would make a global and combined fit of all the diboson channels possible, and we can expect this to further improve the sensitivity to new physics fully exploiting the potential of FCC. It is worth noting that a combination of all channels could also be beneficial in reducing systematic and theory uncertainties, which are partially correlated across different processes.

\section*{Acknowledgments}

We thank S.~Banerjee, R.S.~Gupta and M.~Spannowsky for helpful discussions. We thank J.~De Blas and J.~Gu for providing numerical bounds from a global fit in the Warsaw basis. M.M.~was supported by the Swiss National Science Foundation, under Project Nos.~PP00P2176884. S.D.C., L.D.R.~and G.P.~were supported in part by the MIUR under contract 2017FMJFMW (PRIN2017). F.B.,~P.E.,~C.G.~and A.R.~acknowledge support by the Deutsche
Forschungsgemeinschaft (DFG, German Research Foundation) under Germany’s Excellence
Strategy – EXC 2121 “Quantum Universe” – 390833306. The work of C.G. and A.R. was also supported by the International Helmholtz-Weizmann Research School for Multimessenger Astronomy, largely funded through the Initiative and Networking Fund of the Helmholtz Association. F.B. was also supported by the ERC Starting Grant NewAve (638528). This work was partially performed at the Aspen Center for Physics, which is supported by National Science Foundation grant PHY-1607611.


\newpage

\appendix

\section{Amplitudes}
\label{app:HelAmps}
In this appendix we collect the explicit expressions for the SM and BSM helicity amplitudes. 

\subsection{Helicity amplitudes for $q \bar q \to Zh$}
\label{app:amps_wh}
\newcommand{\swsq}{s_\textsc{w}^2}
\newcommand{\cwsq}{c_\textsc{w}^2}
\newcommand{\sw}{s_\textsc{w}}
\newcommand{\cw}{c_\textsc{w}}
In this section, we report the exact $q \bar q\to Zh$ helicity amplitudes at tree level. For convenience, we define $\varepsilon_Z\equiv \mz/\rtsha$ 
and $\varepsilon_H\equiv\mh/\rtsha$. The scattering angle $\theta$ is defined as in Ref.~\cite{Bishara:2020vix}. The $Z$ boson polarization vectors are defined with respect to the null reference momentum $(|\vec{p}_Z|,\,-\vec{p}_Z)$, where $p_Z$ is the $Z$ momentum in the $Zh$ center of mass frame. 
The helicity of the fermion is denoted by $h=-1 (+1)$ for a left- (right-) handed fermion and its electric charge by $Q$.
The coupling of the SM $Z$-boson to the fermions is denoted by $g_{Z,f}^h$ where $g_{Z,f}^- = (T_{3,f}-\swsq Q_f)/(\sw\cw)$ and $g_{Z,f}^+ = -\sw Q_f/\cw$.
Note that a common phase between any given SM and BSM helicity amplitude is not physical (we stress that it must be a common and not a relative phase) -- all such phases have been removed for clarity.

\begin{equation}
\begin{split}
\mcM_{\text{SM},\pm}&=
-\frac{4\pi\alpha\,g^h_{Z,f}}{\sqrt{2}\cw\sw}
\,
\frac{M_Z}{\sqrt{\hat{s}}}\frac{1 \pm h \cos{\theta}}{1-\varepsilon_{Z}^{2}}\\
\mcM_{\text{SM},0}&= -\,\frac{2\pi\alpha\,g^h_{Z,f}}{\cw\sw}
\, 
\sin\theta\,\frac{1-\varepsilon_H^2+\varepsilon_Z^2 }{1-\varepsilon_Z^2}   
\end{split}\label{eq:amp_SM}
\end{equation}

\begin{equation}
\begin{split}
\mcM_{\varphi q,\pm}^{(3)} &= 
    -\left(-1\right)^{3Q}\,\frac{1-h}{2}\,\sqrt{2}\,\cfqtrip\,
    \frac{M_Z\sqrt{\hat{s}}}{\Lambda^2}\,
    \frac{1 \pm h \cos{\theta}}{1-\varepsilon_{Z}^{2}}\\
\mcM_{\varphi q,0}^{(3)} &= 
    -\left(-1\right)^{3Q}\,\frac{1-h}{2}\cfqtrip\,
    \frac{\hat{s}}{\Lambda^2}\,\sin\theta\,
    \frac{1-\varepsilon_H^2+\varepsilon_Z^2 }{1-\varepsilon_Z^2}   
\end{split}\label{eq:amp_Oq3}
\end{equation}

\begin{equation}
\begin{split}
\mcM_{\varphi q,\pm}^{(1)} &= 
    \frac{1-h}{2}\,\sqrt{2}\,\cfqsing\,
    \frac{M_Z\sqrt{\hat{s}}}{\Lambda^2}\,
    \frac{1 \pm h \cos{\theta}}{1-\varepsilon_{Z}^{2}}\\
\mcM_{\varphi q,0}^{(1)} &= 
    \frac{1-h}{2}\cfqsing\,
    \frac{\hat{s}}{\Lambda^2}\,\sin\theta\,
    \frac{1-\varepsilon_H^2+\varepsilon_Z^2 }{1-\varepsilon_Z^2}   
\end{split}\label{eq:amp_Oq1}
\end{equation}

\begin{equation}
\begin{split}
\mcM_{\varphi u,\pm} &=
    \frac{1+h}{2}\,\sqrt{2}\,\cfu\,
    \frac{M_Z\sqrt{\hat{s}}}{\Lambda^2}\,
    \frac{1 \pm h \cos{\theta}}{1-\varepsilon_{Z}^{2}}\\
\mcM_{\varphi u,0} &= 
    \frac{1+h}{2}\cfu\,
    \frac{\hat{s}}{\Lambda^2}\,\sin\theta\,
    \frac{1-\varepsilon_H^2+\varepsilon_Z^2 }{1-\varepsilon_Z^2}   
\end{split}\label{eq:amp_Ou}
\end{equation}

\begin{equation}
\begin{split}
\mcM_{\varphi d,\pm} &= 
    \frac{1+h}{2}\,\sqrt{2}\,\cfd\,
    \frac{M_Z\sqrt{\hat{s}}}{\Lambda^2}\,
    \frac{1 \pm h \cos{\theta}}{1-\varepsilon_{Z}^{2}}\\
\mcM_{\varphi d,0} &= 
    \frac{1+h}{2}\cfd\,
    \frac{\hat{s}}{\Lambda^2}\,\sin\theta\,
    \frac{1-\varepsilon_H^2+\varepsilon_Z^2 }{1-\varepsilon_Z^2}   
\end{split}\label{eq:amp_Od}
\end{equation}

\subsection{Squared amplitudes and interference terms for $q \bar q \to Zh$}
\label{app:amps_squared}
\newcommand{\dwsq}{\left|\mathcal{D}_W\right|^2}

The squared SM amplitude and the SM-BSM interference terms are given separately below.
For convenience, we define a function that depends on the scattering angle and is common among all squared and interference terms,
\begin{equation}
    f(\hat{s},\theta)=(1-\varepsilon_H^2+\varepsilon_Z^2)^2\sin^2\theta+\frac{2M^2_Z}{\hat{s}}\left(3+\cos2\theta\right)\,.
\end{equation}
With this definition, the squared and interference terms are,
\begin{equation}
\begin{split}
\left|\mathcal{M}_\text{SM}\right|^2 &=
    \left(\frac{2\pi\alpha}{\cw\sw}\right)^2
    \frac{{g^-_{Z,q}}^2+{g^+_{Z,q}}^2}{(1-\varepsilon_Z^2)^2}\,f(\hat{s},\theta)\,,\\
2\Re\mathcal{M}_\text{SM}\mathcal{M}_{\varphi q}^{(3)*} &=
    \frac{4\pi\alpha}{\cw\sw}
    \frac{(-1)^{3Q}g^-_{Z,f}\,\cfqtrip}{(1-\varepsilon_Z^2)^2}
    \frac{\hat{s}}{\Lambda^2}\,f(\hat{s},\theta)\,,\\
2\Re\mathcal{M}_\text{SM}\mathcal{M}_{\varphi q}^{(1)*} &=
    -\frac{4\pi\alpha}{\cw\sw}
    \frac{g^-_{Z,f}\,\cfqsing}{(1-\varepsilon_Z^2)^2}
    \frac{\hat{s}}{\Lambda^2}\,f(\hat{s},\theta)\,,\\
2\Re\mathcal{M}_\text{SM}\mathcal{M}_{\varphi u}^* &=
    -\frac{4\pi\alpha}{\cw\sw}
    \frac{g^+_{Z,u}\,\cfu}{(1-\varepsilon_Z^2)^2}
    \frac{\hat{s}}{\Lambda^2}\,f(\hat{s},\theta)\,,\\
2\Re\mathcal{M}_\text{SM}\mathcal{M}_{\varphi d}^* &=
    -\frac{4\pi\alpha}{\cw\sw}
    \frac{g^+_{Z,d}\,\cfd}{(1-\varepsilon_Z^2)^2}
    \frac{\hat{s}}{\Lambda^2}\,f(\hat{s},\theta)\,.
\end{split}
\end{equation}

\begin{table}
    \centering
    \begin{tabular}{@{\hspace{.3em}}c@{\hspace{.4em}}|@{\hspace{.4em}}c@{\hspace{.4em}}|@{\hspace{.4em}}c@{\hspace{.4em}}|@{\hspace{.4em}}c@{\hspace{.3em}}}\toprule[1.5pt]
         $p_{T \min}$ bin [GeV] &  $Zh\to\ell\ell\gamma\gamma$ & $Zh\to\nu\bar{\nu}\gamma\gamma$ & $Wh\to\nu\ell\gamma\gamma$\\
         \midrule
         \rule{0pt}{1.25em}200 -- 400  & $1+0.52-0.09 = 1.43$ & $1+0.31-0.09 = 1.22$ & $1+0.28-0.08 = 1.20$\\ 
         \rule{0pt}{1.25em}400 -- 600  & $1+0.64-0.14 = 1.50$ & $1+0.37-0.14 = 1.23$ & $1+0.29-0.17 = 1.12$\\ 
         \rule{0pt}{1.25em}600 -- 800  & $1+0.69-0.18 = 1.51$ & $1+0.40-0.18 = 1.22$ & $1+0.36-0.24 = 1.12$\\ 
         \rule{0pt}{1.25em}800 -- 1000 & $1+0.70-0.24 = 1.46$ & $1+0.40-0.24 = 1.16$ & $1+0.37-0.32 = 1.05$\\ 
         \rule{0pt}{1.25em}1000 -- $\infty$ & $1+0.69-0.32 = 1.37$ & $1+0.40-0.32 = 1.08$ & $1+0.37-0.40 =0.97$\\ 
         \bottomrule[1.5pt]
    \end{tabular}
    \caption{NLO k-factors for the main signal processes. Each entry shows separately the QCD and QED contributions to the k-factors. The accuracy on the determination of the k-factors is of order $\textit{few}\times 10^{-2}$.}
    \label{tab:k_factors}
\end{table}

\section{Monte Carlo event generation}\label{app:mc_evt_gen}

The main processes contributing to the signal ($q \bar q \to Zh$ and $q \bar q \to Wh$) were simulated at LO and we took into account QCD and QED NLO effects via k-factors. We computed the QCD k-factors with \textsc{MadGraph5\_aMC@NLO}, while the QED ones were extracted from Ref.~\cite{Frederix:2018nkq}. We verified that, in our bins, NLO QCD corrections have a negligible dependence on the Wilson coefficients, hence we just rescaled the LO cross-sections by the SM k-factors.
The NLO QED k-factors in Ref.~\cite{Frederix:2018nkq} are given
as a function of $p_{T}^{h}$. To compute them in our $p_{T\min}$ bins, we used an event-by-event rescaling.

The k-factors in the various $p_{T,min}$ bins are listed in Table~\ref{tab:k_factors} in the format $1+(k_\textsc{qcd}-1)+(k_\textsc{qed}-1)$. As one can see, they give an enhancement of the cross-section of up to $50\%$.
QCD corrections, which enhance the cross-section, typically dominate. On the other hand, QED effects, which tend to lower the cross-section, are subleading in the low-energy bins, whereas they become comparable to the QCD ones at high energy.

The background processes were simulated at NLO in QCD, with the exception of $gg \to Zh$, which was simulated at LO.  QED corrections have not been applied to the backgrounds. This is a conservative choice since they reduce the cross-sections.

\paragraph{Generation cuts}

We applied a set of cuts at generation level to increase the number of Monte Carlo events after the selection cuts. These generation cuts are shown in Table~\ref{tab:gen_cuts}, where the second (third) column shows the values used for all the processes relevant for the $Z\rightarrow \nu\bar{\nu}$ ($Z\rightarrow \ell^{+}\ell^{-}$) channel. Furthermore, we generated the events in $6$ exclusive bins in the $p_{T}$ of the gauge boson.
We notice that, due to initial state radiation effects, a sizeable migration of events between generation and $p_{T\min}$ bins is present.

\begin{table}[t]
\begin{centering}
\centering\renewcommand*{\arraystretch}{1.5}
\begin{tabular}{c | m{2.5cm} | m{2.5cm} }
\toprule[1.5pt] 
 & \centering{$Z\rightarrow\nu\bar{\nu}$} & \centering{$Z\rightarrow\ell^+ \ell^-$} \tabularnewline
\midrule 
$p_{T,\min}^{l}$ & \centering{$0$} & \centering{$30^{a}$}\tabularnewline
$|\eta_{max}^{l}|$ & \centering{$\infty$} & \centering{$6.1$} \tabularnewline
$p_{T,\min}^{\gamma}$ {[}GeV{]} & \multicolumn{2}{c}{$50^{b}$}\tabularnewline
$|\eta_{max}^{\gamma}|$ & \multicolumn{2}{c}{$6.1^{c}$}\tabularnewline
$\Delta R^{ll,\gamma l}_{\min}$ & \multicolumn{2}{c}{$0.01$}\tabularnewline
$\Delta R^{\gamma\gamma}_{\min}$ & \multicolumn{2}{c}{$0.25^{d}$}\tabularnewline
$p_{T}^{V}$ & \multicolumn{2}{c}{$\{0,\,200,\,400,\,600,\,800,\,1200,\,\infty\}$}\tabularnewline
\bottomrule[1.5pt]
\end{tabular}
\par\end{centering}
\caption{Parton level generation cuts for signal and background processes.
$p_{T}^{V}$ denotes the gauge boson $p_T$.
$^{a}$: only applied to QCD LO runs. $^{b}$: not applied to the photons produced by a Higgs decay. $^{c}$: not applied 
to the photons coming from a Higgs boson. $^{d}$: set to $0.01$ in QCD LO runs.}
\label{tab:gen_cuts}
\end{table}

\paragraph{Fits of the signal and background cross-sections}
In Tables~\ref{tab:sigma_full} and~\ref{tab:sigma_full_lep}, we show the fits of the signal and background cross-section in the various bins as a function of the $c_{\varphi q}^{(3)}$, $c_{\varphi q}^{(1)}$, $c_{\varphi u}$ and $c_{\varphi d}$ Wilson coefficients.

\begin{table}
\begin{center}
\setlength{\extrarowheight}{0mm}%
\scalebox{.86}{
\begin{tabular}{@{\hspace{.15em}}c|c|c|c@{\hspace{.25em}}}
\toprule
\rule[-.5em]{0pt}{.5em}
\multirow{2}{*}{$p_{T,min}$ bin} &\multirow{2}{*}{$|y_{h}|$ bin}& \multicolumn{2}{c}{Number of expected events}\tabularnewline
\cline{3-4} &  & \rule{0pt}{1.15em}Signal & Background \tabularnewline
\midrule
\multirow{2}{*}{\rule{0pt}{3.75em} $[200-400]$\,GeV} & $[0,2]$ &
$\begin{aligned}2574\,& + 21600 \,\cfqtrip
+ 1620\,\cfqsing + 2430 \,\cfu - 1370 \,\cfd\\
& + 61900 \,\left(\cfqtrip\right)^{2}+ 32600 \,\left(\cfqsing\right)^{2}\\
\rule[-.65em]{0pt}{1em}& + 15930\,\left(\cfu\right)^{2} +  16620\,\left(\cfd\right)^{2} + (1500 \pm 600)\,\cfqtrip\,\cfqsing
\end{aligned}$ & $1860$\tabularnewline
\cline{2-4} 
 & $[2,6]$ & $\begin{aligned} \rule{0pt}{1.25em}1928\,& + 15540 \,\cfqtrip
- (610 \pm 32)\,\cfqsing + 2160 \,\cfu - 790 \,\cfd \\
& + 42500 \,\left(\cfqtrip\right)^{2}+ 22600 \,\left(\cfqsing\right)^{2}\\
\rule[-.65em]{0pt}{1em}& + 13240\,\left(\cfu\right)^{2} +  9280\,\left(\cfd\right)^{2} - (7500\pm500)\,\cfqtrip\,\cfqsing
\end{aligned}$ & $1150$ \tabularnewline
\hline 
\multirow{2}{*}{\rule{0pt}{3.75em} $[400-600]$\,GeV} & $[0,2]$ & $\begin{aligned} \rule{0pt}{1.25em}406\,& + 9870 \,\cfqtrip
+ 600\,\cfqsing + 1250 \,\cfu -  620 \,\cfd \\
&+ 79700 \,\left(\cfqtrip\right)^{2}+ 42710 \,\left(\cfqsing\right)^{2}\\
\rule[-.65em]{0pt}{1em}&+ 21870\,\left(\cfu\right)^{2} + 21020\,\left(\cfd\right)^{2} - (1400\pm 400) \,\cfqtrip\,\cfqsing
\end{aligned}$ & $157$ \tabularnewline
\cline{2-4} 
 & $[2,6]$ & $\begin{aligned} \rule{0pt}{1.25em}217\,& + 5050 \,\cfqtrip
- 400\,\cfqsing + 821 \,\cfu - 262 \,\cfd\\
& + 38780 \,\left(\cfqtrip\right)^{2}+ 21950 \,\left(\cfqsing\right)^{2}\\
\rule[-.65em]{0pt}{1em}&+ 13320\,\left(\cfu\right)^{2} + 8550\,\left(\cfd\right)^{2} - 9600\,\cfqtrip\,\cfqsing
\end{aligned}$ & $78\pm6$ \tabularnewline
\hline 
\multirow{2}{*}{\rule{0pt}{3.75em} $[600-800]$\,GeV} & $[0,1.5]$ & $\begin{aligned} \rule{0pt}{1.25em} 75\,& + 3385\,\cfqtrip
+ 215\,\cfqsing + 496\,\cfu - 243 \,\cfd\\
& + 49020 \,\left(\cfqtrip\right)^{2}+ 31340 \,\left(\cfqsing\right)^{2}\\
\rule[-.65em]{0pt}{1em}&+ 16130\,\left(\cfu\right)^{2} + 15200\,\left(\cfd\right)^{2} - (1800\pm 200)\,\cfqtrip\,\cfqsing
\end{aligned}$ & $17\pm1$ \tabularnewline
\cline{2-4} 
 & $[1.5,6]$ & $\begin{aligned} \rule{0pt}{1.25em} 63\,& + 2796 \,\cfqtrip
- 206\,\cfqsing + 487\,\cfu - 169 \,\cfd\\
& + 39130 \,\left(\cfqtrip\right)^{2}+ 25570 \,\left(\cfqsing\right)^{2}\\
\rule[-.65em]{0pt}{1em}&+ 15380\,\left(\cfu\right)^{2} + 10250\,\left(\cfd\right)^{2} - 10100\,\cfqtrip\,\cfqsing
\end{aligned}$ & $17\pm1$ \tabularnewline
\hline 
\multirow{2}{*}{\rule{0pt}{3.75em} $[800-1000]$\,GeV} & $[0,1]$ & $\begin{aligned} \rule{0pt}{1.25em} 10 \,& + 728 \,\cfqtrip
+ (45 \pm 7)\,\cfqsing + 111 \,\cfu - (50 \pm 4) \,\cfd\\
& + 16510 \,\left(\cfqtrip\right)^{2}+ 11320 \,\left(\cfqsing\right)^{2}\\
\rule[-.65em]{0pt}{1em}&+ 5790\,\left(\cfu\right)^{2} + 5460 \,\left(\cfd\right)^{2} - (400\pm 200)\,\cfqtrip\,\cfqsing
\end{aligned}$ & $2.4\pm0.4$ \tabularnewline
\cline{2-4} 
 & $[1,6]$ & $\begin{aligned} \rule{0pt}{1.25em} 16\,& + 1116 \,\cfqtrip - (68 \pm 6)\,\cfqsing + 210 \,\cfu - 71 \,\cfd\\
& + 24920 \,\left(\cfqtrip\right)^{2}+ 17460 \,\left(\cfqsing\right)^{2}\\
\rule[-.65em]{0pt}{1em}&+ 10330\,\left(\cfu\right)^{2} + 7180 \, \left( \cfd \right)^{2} - 6500\,\cfqtrip\,\cfqsing
\end{aligned}$ & $3\pm1$ \tabularnewline
\hline 
\multirow{2}{*}{\rule{0pt}{3.75em}$[1000-\infty]$\,GeV} & $[0,1]$ & $\begin{aligned} \rule{0pt}{1.25em}3\,& + 373 \,\cfqtrip
+ (20 \pm 10)\,\cfqsing + (63 \pm 4) \,\cfu - (20\pm 5) \,\cfd\\
& + 17600 \,\left(\cfqtrip\right)^{2}+ 12550 \,\left(\cfqsing\right)^{2}\\
\rule[-.65em]{0pt}{1em}&+ 6700\,\left(\cfu\right)^{2} + 5800\,\left(\cfd\right)^{2} - (1700\pm 300)\,\cfqtrip\,\cfqsing
\end{aligned}$ & $1.3\pm0.3$ \tabularnewline
\cline{2-4} 
 & $[1,6]$ & $\begin{aligned} \rule{0pt}{1.25em}4\,& + 498 \,\cfqtrip - (30 \pm 10)\,\cfqsing + (88 \pm 5) \,\cfu - (44 \pm 5) \,\cfd\\
& + 22250 \,\left(\cfqtrip\right)^{2}+ 15800 \,\left(\cfqsing\right)^{2}\\
\rule[-.65em]{0pt}{1em}&+ 9670\,\left(\cfu\right)^{2} + 6270\, \left(\cfd\right)^{2} - (7500\pm300)\,\cfqtrip\,\cfqsing
\end{aligned}$ & $1.9\pm0.3$ \tabularnewline
\bottomrule
\end{tabular}}
\end{center}
\caption{Number of expected signal events as a function of the Wilson coefficients (with $\Lambda = 1$ TeV) and background events in the $Z\rightarrow \nu \bar{\nu}$ channel at FCC-hh with $30\,{\rm ab}^{-1}$ integrated luminosity. The Monte Carlo errors on the fitted coefficients, when not explicitly specified, are at most of order $\textit{few}$ \%.
}
\label{tab:sigma_full}
\end{table}

\begin{table}
\begin{center}
\setlength{\extrarowheight}{0mm}%
\scalebox{.86}{
\begin{tabular}{@{\hspace{.15em}}c|c|c|c@{\hspace{.25em}}}
\toprule
\rule[-.5em]{0pt}{.5em}
\multirow{2}{*}{$p_{T,min}$ bin} &\multirow{2}{*}{$|y_{Zh}|$ bin}& \multicolumn{2}{c}{Number of expected events}\tabularnewline
\cline{3-4} &  & Signal & Background \tabularnewline
\midrule
\multirow{2}{*}{\rule{0pt}{3.75em} $[200-400]$\,GeV} & $[0,2]$ &
$\begin{aligned}  433 & + 3330 \,\cfqtrip + 480\,\cfqsing + 651 \,\cfu - 355 \,\cfd\\
& + 8800 \,\left(\cfqtrip\right)^{2}+ 8780 \,\left(\cfqsing\right)^{2}\\
\rule[-.65em]{0pt}{1em}&+ 4270\,\left(\cfu\right)^{2} +  4550\,\left(\cfd\right)^{2} + (600\pm 140)\,\cfqtrip\,\cfqsing
\end{aligned}$ & $339$\tabularnewline
\cline{2-4} 
 & $[2,6]$ & $\begin{aligned} \rule{0pt}{1.25em} 306& + 2270 \,\cfqtrip - (164 \pm 9)\,\cfqsing + 548 \,\cfu - 185 \,\cfd\\
& + 5740 \,\left(\cfqtrip\right)^{2}+ 5680 \,\left(\cfqsing\right)^{2}\\
\rule[-.65em]{0pt}{1em}&+ 3370\,\left(\cfu\right)^{2} + 2320\,\left(\cfd\right)^{2} - (2270\pm130)\,\cfqtrip\,\cfqsing
\end{aligned}$ &  $177$\tabularnewline
\hline 
\multirow{2}{*}{\rule{0pt}{3.75em} $[400-600]$\,GeV} & $[0,2]$ & $\begin{aligned}\rule{0pt}{1.25em} 84& + 1810 \,\cfqtrip
+ 198\,\cfqsing + 360 \,\cfu - 185 \,\cfd\\
& + 12650 \,\left(\cfqtrip\right)^{2}+ 12610 \,\left(\cfqsing\right)^{2}\\
\rule[-.65em]{0pt}{1em}&+ 6340\,\left(\cfu\right)^{2} + 6280\,\left(\cfd\right)^{2} - (240\pm 90)\,\cfqtrip\,\cfqsing
\end{aligned}$ & $31$ \tabularnewline
\cline{2-4} 
 & $[2,6]$ & $\begin{aligned}\rule{0pt}{1.25em} 43 & + 892\,\cfqtrip
- 130\,\cfqsing + 230 \,\cfu - 71 \,\cfd\\
& + 5940 \,\left(\cfqtrip\right)^{2}+ 5980\,\left(\cfqsing\right)^{2}\\
\rule[-.65em]{0pt}{1em}&+ 3710\,\left(\cfu\right)^{2} + 2250\,\left(\cfd\right)^{2} - 2960\,\cfqtrip\,\cfqsing
\end{aligned}$ & $13$ \tabularnewline
\hline 
\multirow{2}{*}{\rule{0pt}{3.75em} $[600-800]$\,GeV} & $[0,2]$ & $\begin{aligned}\rule{0pt}{1.25em} 22& + 936\,\cfqtrip
+ (67\pm 4)\,\cfqsing + 201 \,\cfu - 97\,\cfd\\
& + 12390 \,\left(\cfqtrip\right)^{2}+ 12350 \,\left(\cfqsing\right)^{2}\\
\rule[-.65em]{0pt}{1em}&+ 6430\,\left(\cfu\right)^{2} + 5930\,\left(\cfd\right)^{2} - (950\pm 90)\,\cfqtrip\,\cfqsing
\end{aligned}$ & $5$ \tabularnewline
\cline{2-4} 
 & $[2,6]$ & $\begin{aligned}\rule{0pt}{1.25em} 9& + 346 \,\cfqtrip
- 64\,\cfqsing + 91 \,\cfu - 26 \,\cfd\\
& + 4370 \,\left(\cfqtrip\right)^{2}+ 4380 \,\left(\cfqsing\right)^{2}\\
\rule[-.65em]{0pt}{1em}&+ 2830\,\left(\cfu\right)^{2} + 1560\,\left(\cfd\right)^{2} - 2540\,\cfqtrip\,\cfqsing
\end{aligned}$ & $2$ \tabularnewline
\hline 
\multirow{2}{*}{\rule{0pt}{3.75em} $[800-1000]$\,GeV} & $[0,2]$ & $\begin{aligned}\rule{0pt}{1.25em} 4.7& + 318 \,\cfqtrip
+ (14 \pm 3)\,\cfqsing + 69 \,\cfu - (30 \pm 2) \,\cfd\\
& + 6690 \,\left(\cfqtrip\right)^{2}+ 6720 \,\left(\cfqsing\right)^{2}\\
\rule[-.65em]{0pt}{1em}&+ 3560\,\left(\cfu\right)^{2} + 3130\,\left(\cfd\right)^{2} - (830\pm 70)\,\cfqtrip\,\cfqsing
\end{aligned}$ & $1.1\pm0.08$ \tabularnewline
\cline{2-4} 
 & $[2,6]$ & $\begin{aligned}\rule{0pt}{1.25em} 1.48 & + 95 \,\cfqtrip
- (23 \pm 2)\,\cfqsing + 25 \,\cfu \\
& - (7.6 \pm 0.8) \,\cfd + 1890 \,\left(\cfqtrip\right)^{2}+ 1910 \,\left(\cfqsing\right)^{2}\\
\rule[-.65em]{0pt}{1em}&+ 1252\,\left(\cfu\right)^{2} + 645\,\left(\cfd\right)^{2} - (1240\pm40)\,\cfqtrip\,\cfqsing
\end{aligned}$ & $0.29\pm0.04$ \tabularnewline
\hline 
\multirow{2}{*}{\rule{0pt}{3.75em} $[1000-\infty]$\,GeV} & $[0,2]$ & $\begin{aligned}\rule{0pt}{1.25em} 1.31 & + 163 \,\cfqtrip
- ( 0.4 \pm 4.7) \,\cfqsing + 41 \,\cfu \\
& - ( 14 \pm 2) \,\cfd + 7280 \,\left(\cfqtrip\right)^{2}+ 7310 \,\left(\cfqsing\right)^{2}\\
\rule[-.65em]{0pt}{1em}&+ 4120\,\left(\cfu\right)^{2} + 3240\,\left(\cfd\right)^{2} - (1780\pm 100)\,\cfqtrip\,\cfqsing
\end{aligned}$ & $0.44\pm0.04$ \tabularnewline
\cline{2-4} 
 & $[2,6]$ & $\begin{aligned}\rule{0pt}{1.25em} 0.29 & + 32 \,\cfqtrip - (9 \pm 2)\,\cfqsing + (10.5 \pm 0.9) \,\cfu \\
&- (1.7 \pm 0.8) \,\cfd + 1230 \,\left(\cfqtrip\right)^{2}+ 1240 \,\left(\cfqsing\right)^{2}\\
\rule[-.65em]{0pt}{1em}&+ 842\,\left(\cfu\right)^{2} + 379\,\left(\cfd\right)^{2} - (940\pm40)\,\cfqtrip\,\cfqsing
\end{aligned}$ & $0.12\pm0.03$ \tabularnewline
\bottomrule
\end{tabular}
}
\par\end{center}
\caption{Number of expected signal events as a function of the Wilson coefficients (with $\Lambda = 1$ TeV) and background events in the $Z\rightarrow \ell^+ \ell^-$ channel at FCC-hh with $30\,{\rm ab}^{-1}$ integrated luminosity. The Monte Carlo errors on the fitted coefficients, when not explicitly specified, are at most of order $\textit{few}$ \%.
}\label{tab:sigma_full_lep}
\end{table}

\clearpage
\providecommand{\href}[2]{#2}\begingroup\raggedright\endgroup

\end{document}